\documentclass{article}
\usepackage{arxiv}

\usepackage[utf8]{inputenc} 
\usepackage[T1]{fontenc}    
\usepackage{graphicx}
\usepackage{hyperref}       
\usepackage{url}            
\usepackage{booktabs}       
\usepackage{amsfonts}       
\usepackage{nicefrac}       
\usepackage{microtype}      
\usepackage{natbib}
\usepackage{doi}

\usepackage{mathtools}
\usepackage{amsmath}
\usepackage{amsfonts}
\usepackage{amssymb}
\usepackage{booktabs}       
\usepackage{array}          
\usepackage{subcaption}     
\usepackage{color}

\usepackage{enumitem}       
\usepackage[ruled,vlined,linesnumbered]{algorithm2e}    

\newcolumntype{L}[1]{>{\raggedright\let\newline\\\arraybackslash\hspace{0pt}}m{#1}}
\newcolumntype{C}[1]{>{\centering\let\newline\\\arraybackslash\hspace{0pt}}m{#1}}
\newcolumntype{R}[1]{>{\raggedleft\let\newline\\\arraybackslash\hspace{0pt}}m{#1}}

\newcommand{\normal}[2]{\mathcal{N}(#1, #2)}
\newcommand{\uniform}[2]{\mathcal{U}(#1, #2)}
\newcommand{\uniformset}[1]{\mathcal{U}(#1)}
\newcommand{\bernoulli}[1]{\text{Bern}(#1)}

\newcommand{\norm}[1]{\lVert#1\rVert}		
\newcommand{\vect}[1]{\left[#1\right]}		
\newcommand{\tuple}[1]{\left(#1\right)}		
\newcommand{\card}[1]{\lvert#1\rvert}		

\newcommand{\R}{\mathbb{R}}

\newcommand{\loss}[2][]{\mathcal{L}_{#1}(#2)}

\newcommand{\ipomdp}{\mathcal{M}}           
\newcommand{\interstates}{\check{S}}        
\newcommand{\states}{S}                     
\renewcommand{\models}{M}                   
\newcommand{\model}{m}                      
\newcommand{\st}{s}                         
\newcommand{\actions}{A}                    
\newcommand{\act}{a}                        
\newcommand{\transition}{T}                 
\newcommand{\team}{N}                       
\newcommand{\teamminus}[1]{\team^{-{#1}}}   
\newcommand{\nagents}{N_{\text{ag}}}        
\newcommand{\agents}{\mathcal{A}}           
\newcommand{\observations}{\Omega}          
\newcommand{\obsfunc}{O}                    
\newcommand{\obs}{o}                        
\newcommand{\rwdfunc}{R}                    
\newcommand{\rwdtask}{\rwdfunc_{\text{task}}}   
\newcommand{\features}{\phi}                
\newcommand{\rwdweights}{\theta}            
\newcommand{\policy}{\pi}                   

\newcommand{\profile}{\psi}                 
\newcommand{\profiles}{\Psi}                
\newcommand{\traj}[2][]{\tau_{#2#1}}        
\newcommand{\trajlen}{T}                    

\newcommand{\paramchar}{\theta}                         
\newcommand{\netchar}{f_\paramchar}                     
\newcommand{\parammental}{\phi}                         
\newcommand{\netmental}{g_\parammental}                 
\newcommand{\parampred}{\gamma}                         
\newcommand{\netpred}{d_\parampred}                     
\newcommand{\echar}[1][]{e_{\text{char}#1}}             
\newcommand{\emental}[1][]{e_{\text{ment}#1}}           
\newcommand{\npast}{N_{\text{past}}}                    
\newcommand{\curtrajlen}{\trajlen_{\text{cur}}}         
\newcommand{\pasttrajlen}{\trajlen_{\text{past}}}       

\newcommand{\trajset}{\mathcal{D}}                      
\newcommand{\jointtraj}[1][]{\boldsymbol{\tau}^{#1}}    
\newcommand{\jointobstraj}[1][]{\hat{\boldsymbol{\tau}}^{#1}}    
\newcommand{\diffsteps}{K}                              
\newcommand{\paramdenoise}{\delta}                      
\newcommand{\diffproc}{q}                               
\newcommand{\denoisemodel}{p_{\paramdenoise}}           
\newcommand{\noise}{\epsilon}                           
\newcommand{\noisemodel}{\noise_\paramdenoise}          
\newcommand{\paraminvdyn}{\xi}                          
\newcommand{\invdyn}{I_{\paraminvdyn}}                  
\newcommand{\horizon}{H}                                
\newcommand{\history}{C}                                
\newcommand{\conditions}[1]{y({#1})}                    
\newcommand{\condguide}{\omega}                         
\newcommand{\lossdiff}{\loss[\text{Diff}]{\paramdenoise}}       
\newcommand{\lossinvdyn}{\loss[\text{Dyn}]{\paraminvdyn}}       
\newcommand{\trajplan}{\traj{\text{plan}}}
\newcommand{\obsdiff}{\delta_{\text{obs}}}              
\newcommand{\obsnorm}{\eta}                             
\newcommand{\obsthresh}{\lambda}                        

\newcommand{\eg}{\textit{e.g.},~}
\newcommand{\Eg}{\textit{E.g.},~}
\newcommand{\ie}{\textit{i.e.},~}

\newcommand{\thetitle}{ToMCAT: Theory-of-Mind for Cooperative Agents in Teams via Multiagent Diffusion Policies}
\renewcommand{\shorttitle}{ToMCAT: Theory-of-Mind for Cooperative Agents in Teams via Multiagent Diffusion Policies}
\newcommand{\thekeywords}{Multiagent Learning, Meta-learning, Theory-of-Mind, Diffusion Policies, Adaptive Agents, Team Modeling}
\newcommand{\theauthors}{Pedro Sequeira, Vidyasagar Sadhu and Melinda Gervasio}

\newcommand{\theabstract}{
In this paper we present ToMCAT (Theory-of-Mind for Cooperative Agents in Teams), a new framework for generating ToM-conditioned trajectories. It combines a meta-learning mechanism, that performs ToM reasoning over teammates' underlying goals and future behavior, with a multiagent denoising-diffusion model, that generates plans for an agent and its teammates conditioned on both the agent's goals and its teammates' characteristics, as computed via ToM. We implemented an online planning system that dynamically samples new trajectories (replans) from the diffusion model whenever it detects a divergence between a previously generated plan and the current state of the world. We conducted several experiments using ToMCAT in a simulated cooking domain. Our results highlight the importance of the dynamic replanning mechanism in reducing the usage of resources without sacrificing team performance. We also show that recent observations about the world and teammates' behavior collected by an agent over the course of an episode combined with ToM inferences are crucial to generate team-aware plans for dynamic adaptation to teammates, especially when no prior information is provided about them.
}

\title{\thetitle}

\author{
\theauthors\\
SRI International\\
333 Ravenswood Ave., Menlo Park, 94025 CA\\
\texttt{pedro.sequeira@sri.com, srikanthvidyasagar.sadhu@sri.com, melinda.gervasio@sri.com}
}

\date{}

\renewcommand{\headeright}{}
\renewcommand{\undertitle}{}

\hypersetup{
    pdftitle={\thetitle},
    pdfsubject={\theabstract},
    pdfauthor={\theauthors},
    pdfkeywords={\thekeywords},
}

\begin{document}

\maketitle              

\begin{abstract}
\theabstract
\end{abstract}


\section{Introduction}%
\label{Sec:Intro}

\begin{figure*}[!ht]
    \centering
    \includegraphics[width=0.9\textwidth]{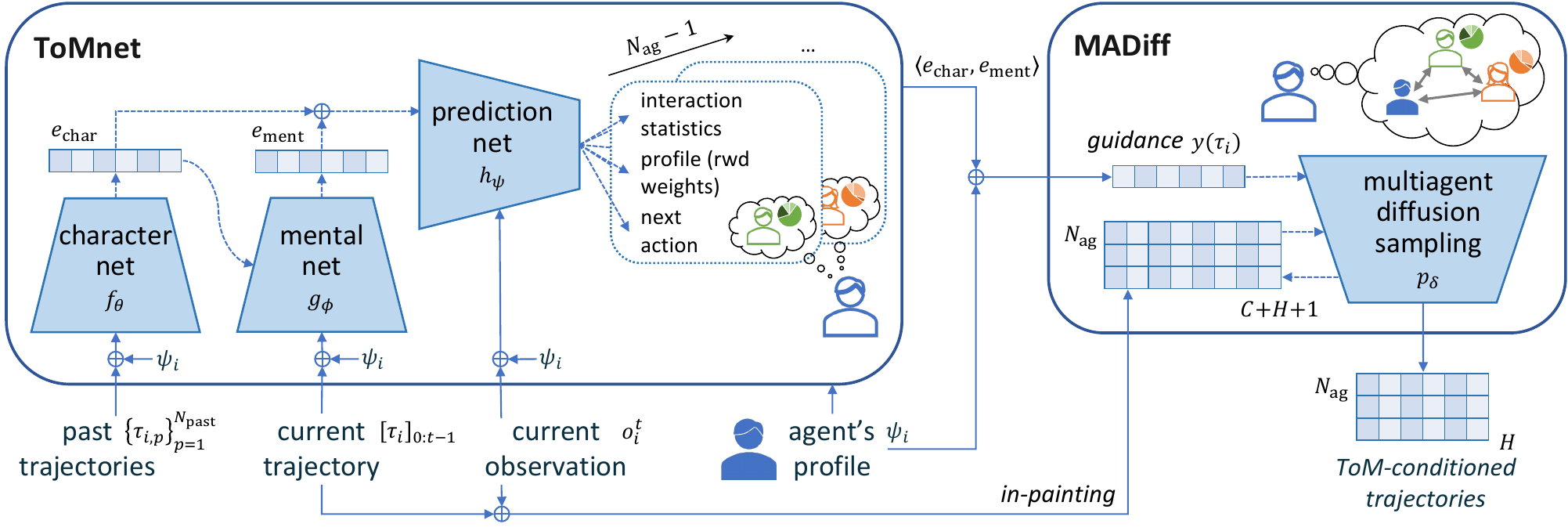}
    \caption{The ToMCAT architecture. Left: a Theory-of-Mind network (ToMnet) makes predictions about the behavior and underlying motivations of the various teammates. Right: a Multiagent Diffusion model (MADiff) generates trajectories conditioned on the ToM reasoning. $\oplus$ indicates a concatenation operation.}
	\label{Fig:Framework}
\end{figure*}

Recent advances in Artificial Intelligence (AI) present an opportunity to greatly improve collaboration between humans and computers to address complex tasks. In particular, AI has the ability to easily generate long courses of action that maximize joint gains, reason about uncertainty, and learn from observation. Notwithstanding, effective human-machine collaboration can only be achieved if artificial agents are endowed with cognitive and social mechanisms, including understanding and predicting the behaviors and preferences of others, communicating intentions, developing shared mental models, and coordinating behavior accordingly \citep{dafoe2020coopai}. In humans, such capability is referred to as forming a \emph{Theory-of-Mind} (ToM) \citep{baroncohen1985tom}.

In this paper, we focus on developing mechanisms that allow understanding other agents'---human or artificial---underlying motivations and predict and adapt to changes in their behavior in order to maximize individual rewards in mixed-motivation settings, all \emph{while} interacting with them. To address this problem, we propose a novel framework called ToMCAT (Theory-of-Mind for Cooperative Agents in Teams).%
\footnote{Throughout the paper we use ``teams'' to refer to \emph{groups of agents} in general. While in our experiments we focus on collaboration, ToMCAT's mechanisms are agnostic to the alignment of the agents' self interests, \ie can be used in collaborative or competitive settings. Similarly, we use ``teammate'' to refer to any other agent in the group.}
Our approach is delineated in Fig.~\ref{Fig:Framework} and comprises two modules: a ToM-predictive network (ToMnet, left) and a multiagent denoising-diffusion policy model (MADiff, right). 

The ToMnet, based on the work of \citet{rabinowitz2018tomnet}, is a neural network that learns a strong prior over agents' preferences and behavior from data generated by some known family of agents. After training, it can be used to make predictions about a particular agent given only limited data about its behavior---a process referred to as meta-learning. We extend the ToMnet framework in \citep{rabinowitz2018tomnet} in various ways. First, we extend it to multiagent settings, where the observer performing the ToM reasoning is itself an active member of a team; therefore predictions are made about multiple teammates, not only over a single agent, and need to be conditioned on the observer's characteristics. In addition, we assume that predictions about teammates are made solely based on the observer's own (partial) current and past observations rather than an external observation function as in \citep{rabinowitz2018tomnet}, and therefore observations include information about all teammates, including the observer itself. Moreover, our ToMnet's outputs represent the observer's beliefs about the mental state of its teammates, and thus \emph{all} predictions are represented as probability distributions.

The MADiff component is based on the work of \citet{zhu2024madiff} and corresponds to a probabilistic denoising-diffusion model that generates multiagent plans. We extend MADiff by guiding the generation of plans based on aspects of both the observer agent and its teammates as encoded in latent representations from the ToMnet, allowing for adaptation via \textit{ToM-conditioned} planning. We also provide a new online mechanism that determines \emph{when} to replan based on a measure of misalignment between planned and real world states, allowing for efficient reuse of MADiff plans.

By combining ToMnet and MADiff, ToMCAT allows an agent not only to plan for itself but also to predict the plans of its teammates from its own observations, leveraging ToM reasoning to compensate for the lack of communication and full observability of teammates' states. Moreover, the conditionally generated behavior through MADiff allows for recursive ToM and nested beliefs, \eg allowing an agent to reason about how the different team members model each other, including the agent itself. Furthermore, conditioning plan generation on ToMnet embeddings that are trained to predict behavior characteristics of different teammates allows interacting with unknown teammates (\ie generalizability) and combining characteristics of teammates from different teams, without the need to retrain either model (\ie compositionality).
%
In summary, our contributions are as follows:
\begin{enumerate}
    \item A meta-learning approach to probabilistic multiagent ToM that takes into account others' preferences and behavior and the observer's own characteristics.
    \item A team behavior predictive denoising-diffusion model that allows generating multiagent plans conditioned on ToM inferences, enabling fast understanding of and adaptation to teammates' preferences and behavior, even in the absence of strong priors.
    \item A dynamic conditional replanning mechanism that allows for efficient online agent adaptation---\ie while interacting with others---by computing discrepancies in planned world states.
\end{enumerate}

The paper is organized as follows. We discuss related work in Sec.~\ref{Sec:Related}, provide technical details on our ToMCAT approach in Sec.~\ref{Sec:ToMCAT}, present experimental results in Sec.~\ref{Sec:Experiments} and draw conclusions and outline future work in Sec.~\ref{Sec:Conclusions}. We include additional implementation details and evaluation results in the Appendix (Sec.~\ref{Sec:Appendix}).

\section{Related Work}%
\label{Sec:Related}



\textbf{Multiagent Reinforcement Learning (MARL)}: To create agents that can adapt to a diverse set of teammates, one option is to use Reinforcement Learning (RL) to train policies that maximize a reward function while treating teammates as an element of the dynamic environment. \Eg in independent Q-learning (IQL)~\citep{lee2022investigation} an agent relies only on its observations without considering the actions of others, resulting in non-stationarity and suboptimal performance. To address this issue, one can instead use multiagent RL (MARL). In particular, centralized training and decentralized execution (CDTE) approaches have been proposed \cite {lowe2017maac} where an agent has access to states and actions of other agents during training, resulting in a centralized value function from which individual policies are derived via a policy gradient algorithm \citep{sutton1999functionapprox}. However, such approaches do not support cases where agents cannot directly observe the internal states of others. 


One approach to overcome that limitation is to use self-play (SP)~\citep{silver2017mastering}, where an agent is trained to play against itself. However, this is mostly applicable to zero-sum settings or homogeneous teams, being hard to adapt to different styles of play or roles in a task. Another approach is to train single-agent policies against a known set of agent models. Namely, in population-based training (PBT)~\citep{jaderberg2019human} one first builds a population of diverse agent models and then trains a policy against agents sampled from the population, improving the population during training, \eg using genetic algorithms. 
While ToMCAT follows a similar approach to PBT by learning from a diverse set of teammates, we 
train a single model to learn the distribution of possible team behaviors in a task such that an agent can be assigned any role and adapt to the behavior of arbitrary teammates, including sub-optimal ones w.r.t. the task.

\textbf{Adhoc teamwork}: Another related body of work is that of ad hoc teamwork~\citep{stone2010adhoc}, where an agent needs to collaborate effectively with other agents without prior coordination or explicit pre-training. Some approaches assume that the task is known beforehand, \eg the ODITS~\citep{gu2022online} framework trains a single-agent RL policy by considering all teammates as part of the environment, learning a teamwork situation encoder whose latent representation conditions the policy for online adaptation, akin to the way we condition diffusion policies based on ToM embeddings. Other approaches try to infer the underlying collaborative task itself, assumed unknown to the ad hoc agent. \Eg the ATPO framework~\citep{ribeiro2023adhoc} casts ad hoc teamwork as identifying the task/teammate from a set of known tasks from histories of observations, then selecting the appropriate action from the corresponding pre-computed task policy. One disadvantage of RL-based approaches is that if an agent's objectives need to change during a task, \eg for role reassignment, we need to re-train the policy, which can be computationally expensive. In contrast, ToMCAT does not train (a set of) RL policies explicitly optimizing for a single reward function---rather, it uses denoising-diffusion policies to generate plans guided by both the agent's own preferences, which might not be aligned with those of its teammates, and teammates' perceived behavior characteristics. In addition, our models make predictions about the underlying attributes of teammates, \ie from a family of parameterized agents, rather than trying to identify a teammate from a discrete set.

\textbf{Theory-of-Mind reasoning}: There is also a wealth of works trying to imbue ToM capabilities in autonomous agents (see, \eg \citep{gurney2022tom,rocha2023tom}). Here we mention those that are mostly related to our approach on conditioning agent behavior on predictions of others' underlying motivations and observed behavior. Some, like ours, adopt the framework of I-POMDPs \citep{gmytrasiewicz2005ipomdp, han2018learning} (an introduction to which is provided in Sec.~\ref{Subsec:Prelim}), where models of others are explicitly added as part of the decision-making space of an agent. \Eg the I-POMDP-Net~\citep{han2019ipomdp} integrates a planning algorithm to approximately solve POMDPs called QMDP~\citep{Karkus2017qmdpnet} and beliefs about models of others into a single neural network for ToM based multi-agent decision making. The PsychSim framework~\citep{pynadath2005psychsim} is a multiagent-based simulation tool based on I-POMDPs for modeling social interaction, where agents form beliefs about the world and maintain recursive models of other agents (ToM). It uses exact Bayesian inference to update the beliefs upon new observations, allowing an agent to compute optimal actions conditioned on the ToM models through forward planning. 
Other approaches employ inverse RL (IRL) to model the motivations of others from observed behavior, which can be seen as a form of ToM \citep{jaraettinger2019tomirl}. \Eg the MIRL-ToM approach in \citep{wu2023multiagent} allows recovering the reward functions used by different agents given trajectories of their behavior by explicitly updating posterior distributions over a set of known models via Bayesian ToM, which allows modeling adaptation to others without requiring a joint equilibrium solution. Overall, these approaches only work with fixed agent profiles and/or a fixed number of teammates, and the (explicit) Bayesian belief update can become computationally intractable as the number of teammates grows. In contrast, ToMCAT does not use an explicit planning module to solve for the I-POMDP and instead learns a distribution of team behaviors from a rich dataset and uses diffusion policies for fast online adaptation to different teammates.

\section{ToMCAT}%
\label{Sec:ToMCAT}

As depicted in Fig.~\ref{Fig:Framework}, our approach uses two models to generate ToM-conditioned behavior for online prediction and adaptation in the context of teams. The Theory-of-Mind network (ToMnet, left) is responsible for making predictions about teammates' behavior conditioned on their observed behavior, while a Multiagent Diffusion network (MADiff, right) generates agent trajectories for all team members conditioned on the ToM reasoning. 

\subsection{Preliminaries}%
\label{Subsec:Prelim}

We model the problem of an agent modeling and adapting to the behavior of others using the formalism of interactive partially observable Markov decision processes (I-POMDPs) \citep{han2019ipomdp}. I-POMDP allows unifying the ToMnet \citep{rabinowitz2018tomnet} and MADiff \citep{zhu2024madiff} approaches---it supports modeling different types or families of agents in the same task as in the ToMnet approach, and characterizing the decision-making problem of individuals in multiagent settings as the MADiff approach. However, it departs from the latter in that it models self-interested agents and is thus applicable to general-sum tasks. Formally, an I-POMDP for an agent $i$ is defined as the tuple $\ipomdp=\tuple{\interstates_i, \actions, \transition_i, \rwdfunc_i,\observations_i,\obsfunc_i}$, where $\interstates_i=\states \times_{j=1}^{\nagents} \models_j$ is the set of interactive states, where $\states$ is the physical environment state space and $\models_j$ is the set of possible models for agent $j$ representing its internal state, $\actions$ is the joint action space, $\rwdfunc_i:\interstates_i\times\actions \to \R$ is agent $i$'s reward function, $\transition_i:\states \times \actions \to \states$ the transition function, $\observations_i$ the observation space, and $\obsfunc_i:\states \times \actions_i \to \observations_i$ the conditional observation function. 

For a given task, we assume a family of agents $\agents=\bigcup_{i} \agents_i$, akin to player \emph{types} in Bayesian game theory \citep{harsanyi1967bayes}, where each agent type is characterized by some private attribute vector, \ie unobservable by others, containing all information relevant to its decision making. Without loss of generality, here we consider individual reward functions of the form $\rwdfunc_i(\st^t)=\boldsymbol{\features}(\st^t)^\top\boldsymbol{\rwdweights}_i$, consisting of linear combinations of (physical) state features, $\features_k:\states \to \R, k=1,\ldots,\card{\boldsymbol{\features}}$, weighted by some weight vector, $\boldsymbol{\rwdweights}_i$. Therefore, we consider that agent types can be succinctly expressed via a \emph{profile} function, $\profile:\agents \to \R^n$, and in our experiments set $\profile_i(\agents_i)\ \coloneq \boldsymbol{\rwdweights}_i$, \ie 
each agent type is motivated to optimize over different aspects of the same general task. For simplicity, we write $\profile_i$ to denote agent $i$'s profile. As for the models $\model_j \in \models_j$ of others, here we allow any representation of teammate $j$'s private information, including predictions over its profile, beliefs or future behavior. 

For training the different modules, we consider observed trajectories of the form $\traj{i}=\{(\obs_i^t,\act_i^t)\}_{t=0}^\trajlen$, where $\obs_i^t = \obsfunc_i(\act_i^{t-1}, \st^t)$ and $\act_i^t \sim \policy_i(\obs_i^t)$ are respectively agent $i$'s observation over the global state, $\st^t$, and its action, taken at trajectory step $t$. We assume that an agent's observations collected through $\obsfunc_i$ include partial information about its teammates' overt behavior that will be used to make predictions about them during task performance.%
\footnote{Such information \emph{does not} include teammates' reward functions, policies or identifiers.}

\subsection{Theory-of-Mind Reasoning}%
\label{Subsec:ToMnet}

As mentioned earlier, our first goal is to create a mechanism capable of inferring the motivations and intent of teammates from their observed behavior. To achieve that, we follow the approach in \citep{rabinowitz2018tomnet}, which aims to develop a machine ToM by training a neural network called a \emph{ToMnet} to model the mental states and behaviors of other agents through meta-learning. The ToMnet learns to predict agents' future actions and infer their goals and beliefs, including false beliefs, by observing their behavior in an environment of interest.

As outlined in Fig.~\ref{Fig:Framework} (left), the ToMnet consists of three modules:
\begin{description}[leftmargin=10pt]
    \item[Character Net:] processes trajectories from past episodes of an observer agent, $i$, acting in the context of a team, with the goal of \emph{characterizing} the general behavioral tendencies of teammates $j\in\teamminus{i}$. Given past trajectories, $\{\traj[,p]{i}\}_{p=1}^{\npast}$, and the observer's profile, $\profile_i$, it outputs a \emph{character embedding}, as $\echar[,i] := \frac{1}{\npast} \sum_{p=1}^{\npast} \netchar(\profile_i, \traj[,p]{i})$,%
    \footnote{In contrast to \citep{rabinowitz2018tomnet}, we average the embedding for each past trajectory when computing $\echar$ to allow for variable number of past trajectories.}
    where $\netchar$ is a learned neural network. The length of past trajectories is denoted by $\pasttrajlen$.
    \item[Mental Net:] focuses on the current episode's trajectory so far to provide the necessary context about teammates' behavior to infer their current \emph{mental} state. It computes a \emph{mental embedding} as $\emental[,i] := \netmental(\profile_i, [\traj{i}]_{t-\curtrajlen:t-1}, \echar[,i])$, where $[\traj{i}]_{\curtrajlen:t-1}$ comprises agent $i$'s last $\curtrajlen$ timesteps in the current trajectory and $\netmental$ is another learned neural network.
    \item[Prediction Net:] uses the outputs of the Character and Mental Nets to update the observer's beliefs about the teammates' behavior and their underlying characteristics. In particular, the Prediction Net is capable of inferring each teammate's next-step action distribution, $\widehat{\policy}_j \coloneq \Delta(\actions_j)$, and the distribution over the sign of its profile elements, $\widehat{\profile_j} \coloneq \Delta(\{-1, 0, 1\}^n)$, given the output of $\netpred(\profile_i, \obs_i^t, \echar[,i], \emental[,i])$, where $\netpred$ is another learned neural network serving as the shared torso for the different prediction heads. The Prediction Net can also estimate other statistics of teammates' future behavior, such as the likelihood of picking up objects in the environment or state occupancies.
\end{description}

The ToMnet is parametrized by $\paramchar$, $\parammental$ and $\parampred$ and is trained end-to-end. It leverages meta-learning to construct a \emph{general ToM}---the model's parameters encode a prior for modeling diverse agent types---while dynamically updating an \emph{agent-specific ToM} --- the model's predictions update a posterior over the characteristics of a particular teammate from recently observed behavior \cite{rabinowitz2018tomnet}. Our extension to multiagent settings results in a single model capable of forming predictions about multiple agent types from the perspective of different types of observer agents. Since others' behavior is contingent on an observer's own behavior, the observer's profile is used as input to condition the predictions of the different modules.

\subsection{Multiagent Diffusion Policies}%
\label{Subsec:MADiff}

The second component of our architecture depicted in Fig.~\ref{Fig:Framework} (right) allows for learning multiagent policies conditioned on the ToM information provided by the ToMnet. The idea is to learn policies that can be used by an agent to adapt to its teammates online, while interacting with them to perform some task. To achieve that, we adopted \emph{Multiagent Diffusion} (MADiff) \citep{zhu2024madiff}, a framework that extends single-agent diffusion policies \citep{janner2022diff,ajay2023condgen} for flexible behavior planning in teams. MADiff trains a conditional generative behavior model from diverse data and treats multiagent planning as sampling from that model. Furthermore, we adopt a \emph{decentralized} execution paradigm, where we assume that each agent makes its own decision without any communication with other agents, based only on local information from its own observations.

Following diffusion probabilistic models \citep{sohldickstein2015diff,ho2020diff}, data generation is modeled with a predefined iterative \emph{diffusion} (forward) process, $\diffproc(\jointtraj[k]|\jointtraj[k-1]) := \normal{\jointtraj[k]; \sqrt{\alpha_k}\jointtraj[k-1]}{(1 - \alpha_k)\boldsymbol{I}}$, and a trainable \emph{denoising} (reverse) process, $\denoisemodel(\jointtraj[k-1]|\jointtraj[k]) := \normal{\jointtraj[k-1]|\mu_\paramdenoise(\jointtraj[k], k)}{\Sigma^k}$, where $k$ is a diffusion/denoising step, $\normal{\mu}{\Sigma}$ denotes a Gaussian distribution with mean $\mu$ and variance $\Sigma$, and $\alpha_k \in \R$ determines the variance schedule.

MADiff is defined as a multiagent offline learning problem, where we have access to a static dataset, $\trajset$, containing joint trajectories of the form $\jointtraj:=\vect{\traj{i}}_{i=0}^{\nagents}$. Since action sequences tend not to be smooth \citep{ajay2023condgen}, we only model observed state sequences as in \citep{zhu2024madiff}. Namely, given a joint trajectory $\jointtraj \in \trajset$, we sample sequences of joint observations, $\jointobstraj:=\vect{(\obs_i^t,\obs_i^{t+1},\ldots,\obs_i^{t+\horizon-1})}_{i=0}^{\nagents}$, where $t \sim \uniform{1}{\trajlen}$ is the time at which an observation was made by agent $i$, sampled uniformly at random, and $\horizon$ is the diffusion horizon. To retrieve actions from generated plans we use $\act_i^t:=\invdyn(\obs_i^t,\obs_i^{t+1})$, where $\invdyn$ is an inverse dynamics model shared amongst agents that is independent from but simultaneously trained with $\denoisemodel$.%

To condition trajectory generation on ToM reasoning, we sample from perturbed distributions of the form $\hat{\denoisemodel}(\jointobstraj) \propto \denoisemodel(\jointobstraj) \conditions{\jointtraj}$, where $\conditions{\jointtraj}$ is an additional input to the diffusion model encoding particular properties of joint trajectory $\jointtraj$. Following the decentralized approach, in ToMCAT perturbations are made from the perspective of an observer agent $i$, which we denote using $\conditions{\traj{i}}$, and can include any of the following information:
\begin{description}[leftmargin=10pt]
    \item[Profile:] the observer agent's profile, $\profile_i$ for which the joint trajectories are being generated.
    \item[Character:] the embedding provided by the Character Net characterizing teammates $j \in \teamminus{i}$, \ie $\echar[,i]\left(\{\traj[,p]{i}\}_{p=1}^{\npast}\right)$, where $\traj[,p]{i} \in \trajset$ are \emph{other} trajectories sampled from the dataset in which agent $i$ interacted with the \emph{same} (types of) teammates.
    \item[Mental:] the Mental embedding characterizing teammates' mental states, \ie $\emental[,i]([\traj{i}]_{0:t-1})$.
    \item[Returns:] any measure of behavior optimality that we wish to condition trajectory generation on, \eg the agent's individual reward, $\rwdfunc_i(\traj{i})$, or the task reward common to all agents. 
\end{description}
%
In situations where we want to condition trajectory generation on multiple factors, we concatenate them to form $\conditions{\traj{i}}$. In addition, we follow \citep{zhu2024madiff,ajay2023condgen,janner2022diff} and constrain trajectory generation on the agent's current observation, $\obs_i^t$, and on the $\history$ previous steps, in a process akin to pixel \emph{in-painting} in image generation \citep{sohldickstein2015diff}. This leads to an augmented trajectory $\jointobstraj:=\vect{(\obs_i^{t-\history}, \ldots, \obs_i^t,\obs_i^{t+1},\ldots,\obs_i^{t+\horizon})}_{i=0}^{\nagents}$.

The MADiff model is parametrized by $\paramdenoise$ and $\paraminvdyn$ which are simultaneously trained by sampling trajectories from dataset $\trajset$ and randomly selecting an agent to be the ``observer'', while other agents are the teammates. Since we follow a decentralized paradigm, the aforementioned in-painting information for the teammates is zeroed-out since the observer does not have access to their observations. Notwithstanding, MADiff is trained to predict the future observations of \emph{all} agents given its local observations.  

\subsection{Dynamic Replanning}%
\label{Subsec:Replanning}

\begin{figure}[ht]
    \centering
    \begin{minipage}{.9\linewidth}
    \begin{algorithm}[H]
    \caption{Online Dynamic Conditional Replanning}
    \label{Alg:Planning}
    \KwIn{$\team$, $i$, $\profile_i$, $\trajset$, $\npast$, $\netchar$, $\netmental$, $\obsnorm$, $\obsthresh$, $\noisemodel$, $\invdyn$, $\condguide$,  $\history$, $\horizon$}
    Initialize 
    $\echar \leftarrow \echar[,i]\left(\{\traj[,p]{i} \sim \trajset \}_{p=1}^{\npast}\right)$, 
    $t \leftarrow 0$, 
    $\hat{\obs_i} \leftarrow \varnothing$, 
    $h\leftarrow \texttt{Queue}(\history, \varnothing)$, 
    $\trajplan \leftarrow \texttt{Queue}(\horizon, \varnothing)$\;
    
    \While{\textnormal{not done}}{
        Agent $i$ observes $\obs_i$; 
        $\texttt{enqueue}(h,\obs_i)$\;
        
        \If{$\textnormal{empty}(\traj{i}) \ || \ \obsdiff(\obs_i,\hat{\obs_i}) > \obsthresh$}{
            $\emental \leftarrow \emental[,i](h)$; 
            $\conditions{\traj{i}} \leftarrow [\echar,\emental,\profile_i]$\;
            
            Initialize $\jointobstraj[\diffsteps] \sim \normal{\boldsymbol{0}}{\alpha\boldsymbol{I}}$\;
            \For{$k=K \ldots 1$}{
                $\jointobstraj[k]_i[:\history+1] \leftarrow h$\;
                $\hat{\noise} \leftarrow \noisemodel(\jointobstraj[k], k) + \condguide \left(\noisemodel(\jointobstraj[k], \conditions{\traj{i}}, k) - \noisemodel(\jointobstraj[k], k) \right)$\;
                
                $\jointobstraj[k-1] \leftarrow \texttt{denoise}(\jointobstraj[k],\hat{\noise}, k)$\;
            }
            $\hat{\traj{}}_i \leftarrow \jointobstraj[0]_i[C+1:]$; 
            $\trajplan \leftarrow \texttt{Queue}(\horizon, \hat{\traj{}}_i)$;
            $\hat{\obs}_i \leftarrow \obs_i$\;
        }
        
        $\hat{\obs}'_i \leftarrow \texttt{dequeue}(\traj{i})$; 
        $\act_i=\invdyn(\hat{\obs}_i,\hat{\obs}'_i)$; 
        $\hat{\obs}_i \leftarrow \hat{\obs}'_i$\;
        Execute $\act_i$, let teammates $j\in \teamminus{i}$ act\;
        
    }
    \end{algorithm}
    \end{minipage}
\end{figure}

To generate plans (trajectories), we use classifier-free guidance \citep{ho2022guidance} with low temperature sampling, where we assume an online setting where an agent $i$ is both observing and interacting with teammates in the context of some task. Generating plans using diffusion models can be computationally intensive \citep{janner2022diff,ajay2023condgen}, so \citet{janner2022diff} proposed a warm-start solution where a plan generated for previous timesteps is reused by first diffusing it for a small number of steps---a fraction of the original diffusion steps---and then denoising the resulting noise for the same number of steps, thereby generating a new plan. However, this approach assumes that plans taken at consecutive timesteps are consistent with one another. Because ToMCAT generates multiagent plans in a decentralized manner, the likelihood that plans remain consistent in consecutive steps is much lower since it would require teammates' actions to be consistent with the predicted trajectory.

As such, we propose a \emph{dynamic replanning} approach where an agent follows the actions in a generated plan while its predictions, including about its teammates, remain consistent, and replans whenever they are inconsistent or the plan is depleted. Moreover, to reduce the number of denoising steps, we use DDIM sampling \citep{song2021ddim}. This results in the online planning (sampling) approach delineated in Algorithm~\ref{Alg:Planning}, corresponding to a modified version of the algorithm in \citep{ajay2023condgen}. In addition to computing and using $\echar$, $\emental$ and $\profile_i$ as conditioning variables, the other novelty resides in line 4, where we compute the difference between the current observation and the observation predicted in the plan, using a suitable metric, $\obsdiff$. If the observation difference is higher than a predefined threshold, $\obsthresh$, we generate a new plan, otherwise we continue using the actions from the previously computed one. See Sec.~\ref{Subsec:AppDynReplan} of the Appendix for a detailed description of the algorithm.

\section{Experiments \& Results}%
\label{Sec:Experiments}

In this section, we detail the experiments carried out to assess the usefulness of ToMCAT in generating dynamic, ToM-conditioned strategies for online adaptation to diverse teammates in collaborative tasks. We detail our experimental scenario, how we trained various RL agents to generate ground-truth behavioral data, and how we trained ToMnet and MADiff models therefrom.%
\footnote{Implementation and training details are provided in the Appendix.}
We then detail the different experiments and analyze the main results.

\subsection{Data Collection}%
\label{Subsec:Data}

\begin{figure}[!tb]
    \centering
    \includegraphics[width=0.4\columnwidth]{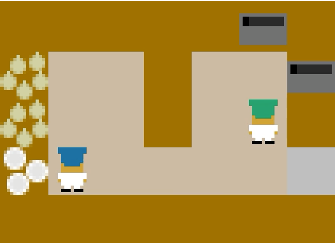}
    \caption{The cooking domain used in the experiments.}
	\label{Fig:Overcooked}
\end{figure}

To test our approach, we designed a scenario in the \emph{Overcooked-AI} simulation platform \citep{carroll2019overcooked},%
\footnote{\url{https://github.com/HumanCompatibleAI/overcooked_ai}}
which is a benchmark domain for research in human-AI collaboration. The custom scenario used in our experiments is illustrated in Fig.~\ref{Fig:Overcooked}. Briefly, the task objective is for a team of $\nagents=2$ agents to collaborate in order to maximize the number of soups served within a time limit. Agents need to collect onions, place them in pots, wait for soups to cook, then serve them in the corresponding station. Each agent has $\card{\actions_i}=6$ available actions at each timestep: they can move in four cardinal directions, perform a ``no-op'' action, or ``interact'' with an object in the environment (\eg pickup or place items). Agents' initial locations in the environment are randomized. Our custom scenario, inspired by the environment in \citep{wu2021cooks}, adds a bottleneck (middle counter) to increase the usefulness of specialized roles in the task and decrease the likelihood that agents can perform the task on their own, requiring effective coordination for success.

As mentioned earlier, here we consider reward functions consisting of linear combinations of state features, and set the profile for an agent $i$ as $\profile_i \coloneq \boldsymbol{\rwdweights}_i$, where $\boldsymbol{\rwdweights}_i$ is the vector of weights associated with each feature. We designed a set of representative profiles capturing varied sub-goals in the Overcooked task (see Table~\ref{Table:Profiles} in the Appendix the corresponding weights, $\boldsymbol{\rwdweights}_i$, and in Sec.~\ref{Subsec:AppProfiles} a description of each reward feature). The profiles range from specialized roles, \ie Cook and Server, to agents that are not task-oriented, \ie Random and Follower, where the latter can result in antagonistic behavior, \eg they can unintentionally block the other agent. Our goal was to design profiles that lead to distinct behaviors in the task and require different types of collaboration.%
\footnote{Although we design a small set of representative profiles, the features form a space of possible profiles, corresponding to the admissible family of agents in the task, $\agents$.}

\begin{figure*}[!tb]
    \centering
    \begin{subfigure}[b]{0.35\textwidth}
        \includegraphics[height=140pt]{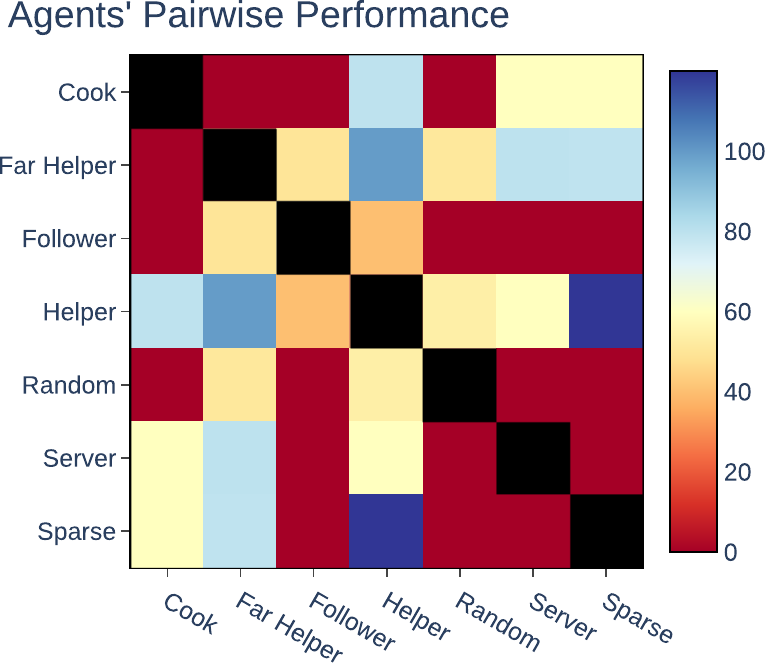}
        \caption{MARL Pairwise Performance}%
        \label{Fig:ResMARLPairwise}
    \end{subfigure}
    \begin{subfigure}[b]{0.4\textwidth}
        \includegraphics[height=140pt]{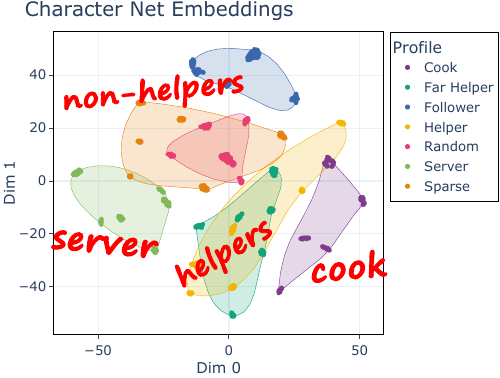}
        \caption{Character Embeddings}%
        \label{Fig:ResTomnetCharEmb}
    \end{subfigure}
    \begin{subfigure}[b]{0.25\textwidth}
        \includegraphics[height=140pt]{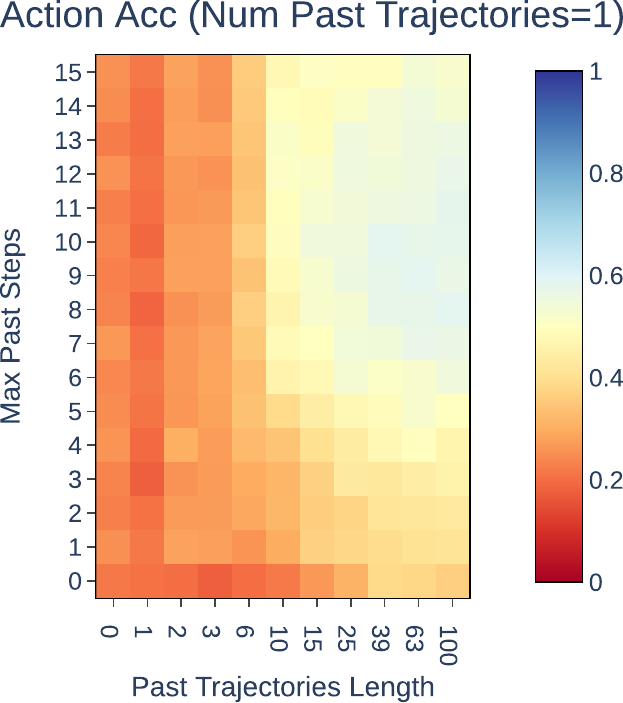}
        \caption{ToMnet Action Accuracy}%
        \label{Fig:ResTomnetActAcc}
    \end{subfigure}
    \caption{MARL and ToMnet training results.}
	\label{Fig:ResultsDataTomnet}
\end{figure*}

We wanted the dataset $\trajset$ used to train our ToM and planning modules to be diverse, showing different ways of satisfying combinations of conditions, $\conditions{\traj{i}}$, so we trained decentralized MARL policies for each pair of profiles in Table~\ref{Table:Profiles} using PPO~\citep{schulman2017ppo} (see details in Sec.~\ref{Subsec:AppMARL} of the Appendix), resulting in $21$ teams.
We then used the trained policies to generated the dataset, $\trajset$, by rolling out trajectories $1\,000$ trajectories for each agent pair, with a maximum trajectory length of $\trajlen=200$ timesteps and a policy exploration (softmax) temperature of $0.3$ to ensure some stochasticity in agents' action selection, resulting in $\card{\trajset}=21\,000$ joint trajectories. 
Fig.~\ref{Fig:ResMARLPairwise} shows the pairwise performance of the different agent pairs in the dataset, as measured by the mean task reward, $\rwdtask$, corresponding to a reward of $\rwdtask=20$ when a soup delivered, and $0$ otherwise. As we can see, the dataset represents a wide range of different team behavior, as desired for training our predictive models. It also shows the impact of pairing an agent trained to optimize for some reward function (\ie profile) with different types of teammates, thus showing inter-dependencies in the task leading to varying degrees of performance (see, \eg results for the Sparse agent).

\subsection{Model Training}%
\label{Subsec:Training}

We trained the ToMnet in a supervised manner, where we first sampled a profile, $\profile_i \sim \profiles$, where $\profiles$ is the set of possible profiles listed in Table~\ref{Table:Profiles}. We set agent $i$ to be the observer, and sampled a joint trajectory, $\traj{} \sim \trajset$ where the team, $N \coloneq \{i,j\}$, is composed of the observer and a teammate, $j$. We then selected a timestep, $t \sim \uniform{1}{\trajlen}$, select the ``current'' observation, $\obs_i^t$, and set the ``current'' trajectory to $[\traj{i}]_{t-\curtrajlen:t-1}$, where $\curtrajlen=10$. We further selected $\npast=4$ past trajectories, $\{\traj[,p]{i} \sim \trajset \}_{p=1}^{\npast}$, that capture prior information about the teammate, \ie where each $\traj[,p]{i}$ contains data of the same team, $\team$. All these data were used as inputs to the ToMnet. Different statistics were then computed from $[\traj{i}]_{t+1:\trajlen}$, which along with the sign of each of the teammate's profile weights, $\boldsymbol{\rwdweights}_j$, and its next action, $\act_j^t$, formed the targets of the Prediction net. See Sec.~\ref{Subsec:AppToMNET} of the Appendix for details.

To train the MADiff component, we sampled joint trajectories $\traj{} \sim \trajset$ similarly to how we trained the ToMnet, but augmented trajectories by computing the ToM embeddings, $\echar[,i]$ and $\emental[,i]$, for each timestep using the trained ToMnet, so they could be used as conditioning variables in $\conditions{\traj{i}}$. In addition, when sampling a trajectory for training, we used $\history=16$ steps to constrain trajectory generation using in-painting, and the subsequent $\horizon=64$ steps as the planning horizon. As explained earlier, here we consider decentralized execution, where observers have access to their local observations, but generate predictions for the future behavior of the whole team. To ensure that, we masked out teammate observations when using history in-painting. After training, we used DDIM sampling for $k=15$ diffusion steps to generate plans in all experiments. See Sec.~\ref{Subsec:AppMADiff} of the Appendix for more details. 

\subsection{Model Analysis}%
\label{Subsec:Analysis}

We now provide some insights from training the ToMnet and MADiff models. Regarding the ToMnet, Fig.~\ref{Fig:ResTomnetCharEmb} plots the 2D t-SNE representations of the character embeddings, $\echar$, computed for the trajectories in $\trajset$. We can see that the Character Net learns different internal representations for different profile ``families'' denoting distinct behavior tendencies in the task. As in \citep{rabinowitz2018tomnet}, this shows that the ToMnet learns useful representations about teammates' characteristics. However, because we condition ToMnet predictions on the observer's own profile, our results further show that it distinguishes between different types of interacting partners, who may require different types of collaboration.
Fig.~\ref{Fig:ResTomnetActAcc} plots the prediction accuracy of the teammate's next action when varying the length of past trajectories, $\pasttrajlen$ and the number of previous steps in the current trajectory $\curtrajlen$, provided to the ToMnet, while fixing $\npast=1$. We observe that with a prior trajectory of only approximately $\pasttrajlen=40$ steps and information on the last $\curtrajlen=10$ steps, the ToMnet achieves maximal performance. The main insight is that the ToMnet does not require much past information about a teammate in order to make accurate predictions about its underlying behavior characteristics. This is an indication that, even in the absence of prior data about a teammate, an observer agent could potentially use observations collected during a trajectory and feed those as past trajectories to compute character embeddings.

\begin{table*}[!tb]
    \centering
    \caption{Effects of conditioning variables, $\conditions{\traj{i}}$, on MADiff training. We also compare the use of history ($\history=16$) vs. no history ($\history=0$) for in-painting constraining. Reported values correspond to averages of cumulative task and individual rewards over $500$ episodes of length $200$ timesteps. Errors correspond to the $95\%$ CI for the mean.}
    \label{Table:MADiffConditioning}
    \begin{tabular}{l r @{\hspace{2pt}$\pm$\hspace{2pt}} r r @{\hspace{2pt}$\pm$\hspace{2pt}} r r @{\hspace{2pt}$\pm$\hspace{2pt}} r r @{\hspace{2pt}$\pm$\hspace{2pt}} r}
        \toprule
        & \multicolumn{4}{c}{\textbf{Task Reward}}
        & \multicolumn{4}{c}{\textbf{Individual Reward}} \\
        
        \multicolumn{1}{l}{\textbf{Conditioning}, $\boldsymbol{\conditions{\traj{i}}}$}
        & \multicolumn{2}{c}{\textbf{No History}}
        & \multicolumn{2}{c}{\textbf{History}}
        & \multicolumn{2}{c}{\textbf{No History}}
        & \multicolumn{2}{c}{\textbf{History}} \\
        \midrule

        $\left[\ \right]$ \emph{(unconditioned)} & $6.68$ & $0.82$ & $9.44$ & $1.44$ & $-54.33$ & $0.58$ & $-52.98$ & $0.76$ \\
        $\left[ \rwdtask \right]$ & $\boldsymbol{19.36}$ & $\boldsymbol{1.30}$ & $\boldsymbol{34.60}$ & $\boldsymbol{1.61}$ & $-85.57$ & $1.05$ & $-87.84$ & $0.78$ \\
        $\left[ \profile_i \right]$ & $4.68$ & $0.83$ & $9.16$ & $1.28$ & $-16.83$ & $0.29$ & $-33.19$ & $0.39$ \\
        $\left[ \echar \right]$ & $6.20$ & $1.09$ & $18.92$ & $1.77$ & $-11.88$ & $0.25$ & $-7.90$ & $0.34$ \\
        $\left[ \emental \right]$ & $9.48$ & $1.31$ & $12.12$ & $1.51$ & $-13.24$ & $0.30$ & $-28.03$ & $0.43$ \\
        $\left[ \echar, \emental \right]$ & $10.04$ & $1.36$ & $22.76$ & $1.89$ & $\boldsymbol{-9.44}$ & $\boldsymbol{0.24}$ & $-2.52$ & $0.31$ \\
        $\left[ \profile_i, \echar, \emental \right]$ & $8.20$ & $1.28$ & $24.88$ & $1.92$ & $-10.67$ & $0.29$ & $\boldsymbol{1.66}$ & $\boldsymbol{0.30}$ \\
        $\left[\rwdtask, \profile_i, \echar, \emental \right]$ & $7.60$ & $0.95$ & $19.20$ & $1.52$ & $-19.63$ & $0.35$ & $-11.05$ & $0.34$ \\

        \bottomrule
    \end{tabular}
\end{table*}

Regarding the MADiff model, we studied the impact of the different ToM conditioning variables and history in-painting on both task and individual performance, which is measured as the cumulative task reward, $\rwdtask$, and individual reward, $\rwdfunc_i$, received over the course of an episode. The latter is important because it is a measure of how well an agent adheres to the role it was assigned for the task. Table~\ref{Table:MADiffConditioning} shows the mean performance for the different types of conditioning, $\conditions{\traj{i}}$, when models were trained with ($\history=16$) and without ($\history=0$) history in-painting. Episodes were generated by having two ToMCAT agents, equipped with the same MADiff model but different profiles, independently generate a plan at each time step and perform the first action of the plan in the environment. 

In general, we observe that the history information provides the necessary context for prediction in models conditioned by ToM-based variables, \ie $\echar$, $\emental$ and $\profile_i$, leading to significantly better performance compared to the no-history counterparts. Furthermore, by looking at the different conditioning sets for the history case, we see that the more ToM ``features'' are used, the better the performance is. We note that even though conditioning only on task rewards, \ie $\rwdtask$, attains the highest team performance, it leads to very poor individual performance. We hypothesize that because the MADiff model is unaware of individual preferences, it learns to mimic the behavior of the best team, leading to improved task performance but failing to carry out its individual responsibilities. In addition, as will be shown in our experiments, conditioning on returns works only if \emph{both} agents use the same MADiff model. 

\subsection{Agent Experiments}%
\label{Subsec:AgentExp}

We now address different research questions to assess the usefulness of online dynamic replanning and the extent to which ToMCAT agents, equipped with different ToM capabilities, can adapt to teammates, both in the presence and absence of priors about their behavioral characteristics. For all experiments reported below and for each test condition, we generated $500$ episodes of $200$ timesteps each, where we paired a ToMCAT agent with a RL agent, randomly assigning profiles to each agent at the start of an episode. In addition, in each episode, the policy of the RL agent was selected from the set trained via MARL, as detailed in Sec.~\ref{Subsec:Data}, corresponding to the policy that was concurrently trained with another RL agent using the same profile as the ToMCAT agent. In practice, this means that the RL agent ``knows'' the profile of the ToMCAT agent and selects the corresponding optimal policy.

\subsubsection{Dynamic Replanning}%

\begin{figure*}[!tb]
    \centering
    \begin{subfigure}[b]{0.3\textwidth}
        \includegraphics[height=140pt]{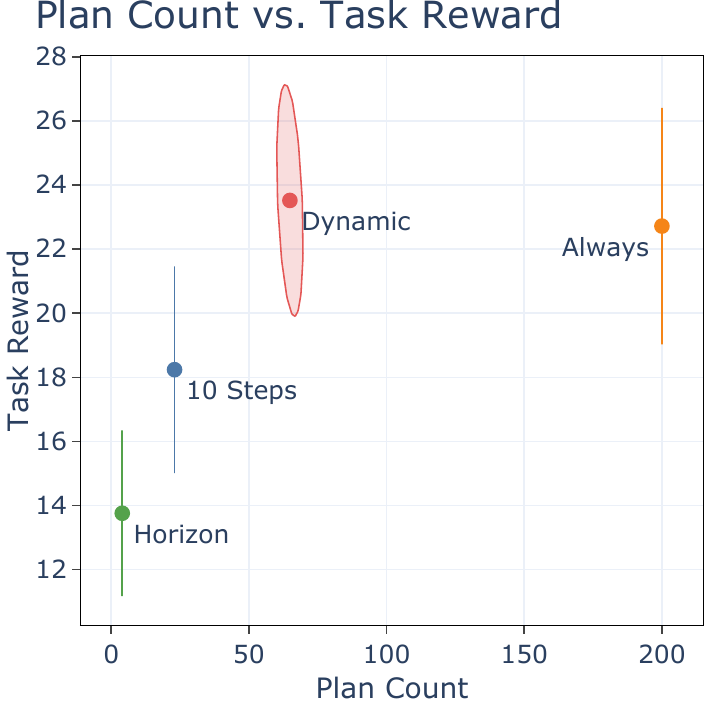}
        \caption{Task Reward}%
        \label{Fig:ResReplanTaskReward}
    \end{subfigure}\hspace{5pt}%
    \begin{subfigure}[b]{0.3\textwidth}
        \includegraphics[height=140pt]{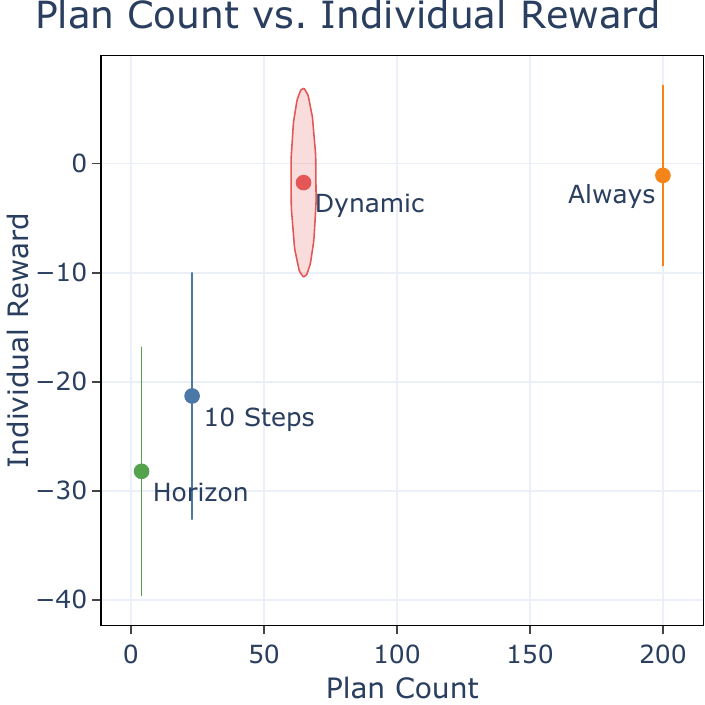}
        \caption{Individual Reward}%
        \label{Fig:ResReplanIndivReward}
    \end{subfigure}\hspace{5pt}%
    \begin{subfigure}[b]{0.3\textwidth}
        \includegraphics[height=140pt]{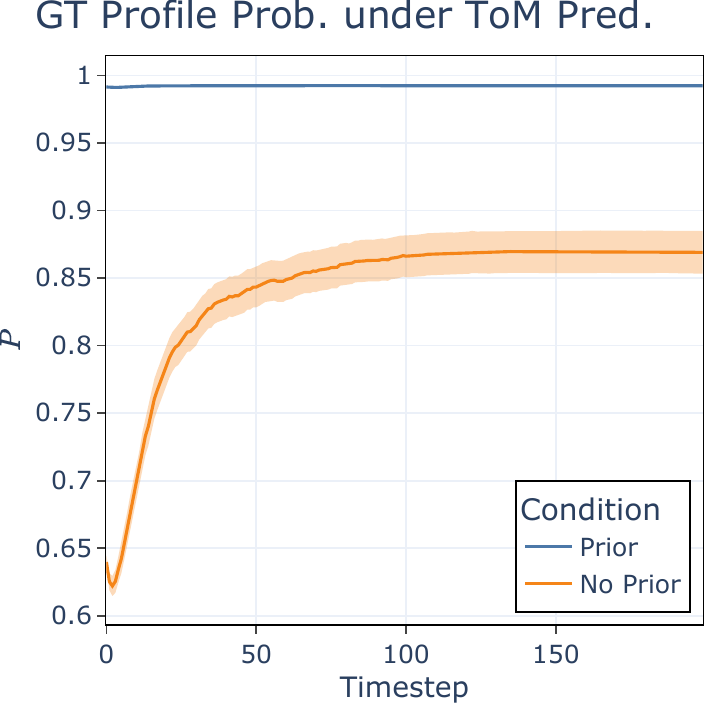}
        \caption{Profile Probability}%
        \label{Fig:ResNoPriorProfProb}
    \end{subfigure}
    \caption{(a)--(b): impact of various replanning schemes on the agents' task and individual cumulative rewards. Covariance ellipses represent the $95\%$ CI of the mean. (c): mean probability over the course of trials of the ground-truth profile of the teammate under the ToMnet prediction in the presence (\textit{Prior}) vs. absence (\textit{No Prior}) of prior information about the teammate.}
	\label{Fig:ResultsReplanning}
\end{figure*}

We want to address the following research question: \textit{``Can the ToMCAT system perform well when deployed in a dynamic environment without requiring replanning at every step?''} To address that question, unlike prior works that measure CPU time on a specific platform, here we use the number of planning steps as a measure of computational ``resource usage'' and assess its impact on task and individual agent performance. 

In this experiment, we used the MADiff model trained with history (\ie $\history=16$) conditioned on all the ToM variables (\ie $\conditions{\traj{i}} \coloneq \left[ \profile_i, \echar, \emental \right]$) as the base model for the ToMCAT agent. To test the usefulness of the online dynamic replanning capability detailed in Sec.~\ref{Subsec:Replanning}, we compare it against replanning at fixed time intervals, namely: at every timestep (\textit{Always}), every $10$ timesteps (\textit{10 Steps}), and every $H=64$ timesteps (\textit{Horizon}). In order to compute the character embedding, $\echar$, the ToMCAT agent is provided $\npast=4$ past trajectories sampled from $\trajset$, where an agent with the same profile interacted with an agent with the teammate's profile.

Figs.~\ref{Fig:ResReplanTaskReward} and \ref{Fig:ResReplanIndivReward} show the average task and individual performance compared to the plan count for different replanning schemes (see Table~\ref{Table:ExpReplanning} of the Appendix for the corresponding numeric data). As we can see, there is a trade-off between task performance and resource usage, and this trade-off is not linear. In particular, \textit{Dynamic} replanning makes the best use of resources compared to fixed-interval schemes, achieving approximately the same level of performance as agents that replan at every step (\ie \textit{Always}), but using only about one third of resources. This shows the advantage of our replanning mechanism that automatically detects the divergence between projected and actual states of the world and replans accordingly, using the entire plan if everything goes as expected.

\subsubsection{Known Teammates}%

In this experiment, we want to address the question: \textit{``Can ToM reasoning improve team performance in the presence of strong teammate priors?''} Here we want to understand whether ToM is important for ToMCAT agents when paired with teammates for whom prior behavior demonstrations are provided, corresponding to supplying $\npast=4$ past trajectories as detailed in the previous experiment. To address that question, we test ToMCAT agents equipped with MADiff models under different conditioning variables, $\conditions{\traj{i}}$, namely: \textit{Unconditioned},  \textit{Returns} ($[ \rwdtask ]$), \textit{Profile} ($[ \profile_i ]$), and \textit{ToM} ($[ \profile_i, \echar, \emental ]$). As an upper bound on performance, we also test against a team of RL agents, where at each episode we sample a pair of policies that were concurrently trained via MARL, corresponding to the \textit{Know RL} condition.

\begin{table}[!t]
    \centering
    \caption{ToM impact in the presence of teammate priors.}
    \label{Table:ExpKnown}
    \begin{tabular}{l r @{\hspace{2pt}$\pm$\hspace{2pt}} r r @{\hspace{2pt}$\pm$\hspace{2pt}} r r @{\hspace{2pt}$\pm$\hspace{2pt}} r}
        \toprule
        \multicolumn{1}{l}{\textbf{Condition}}
        & \multicolumn{2}{c}{\textbf{Plan Count}}
        & \multicolumn{2}{c}{\textbf{Task Rwd}}
        & \multicolumn{2}{c}{\textbf{Indiv. Rwd}} \\
        \midrule

        \emph{Unconditioned} & $78.63$ & $3.62$ & $10.88$ & $1.92$ & $-30.39$ & $8.83$ \\
        \emph{Returns} & $99.71$ & $3.14$ & $11.36$ & $1.56$ & $-37.54$ & $9.75$ \\
        \emph{Profile} & $87.28$ & $3.99$ & $14.24$ & $2.28$ & $-29.29$ & $9.57$ \\
        \emph{ToM} & $\boldsymbol{64.89}$ & $\boldsymbol{3.82}$ & $23.52$ & $2.89$ & $-1.74$ & $6.90$ \\
        \emph{RL Known} & \multicolumn{2}{c}{-} & $\boldsymbol{41.96}$ & $\boldsymbol{3.47}$ & $\boldsymbol{23.85}$ & $\boldsymbol{6.71}$ \\

        \bottomrule
    \end{tabular}
\end{table}

The results are shown in Table~\ref{Table:ExpKnown}, where we measure resource usage (\ie plan count) and the task and individual performance as in the previous experiment. Overall, we observe that agents with ToM capabilities (\ie \textit{ToM)} attain better task performance, adhere the best to their pre-assigned roles (compare with \textit{Profile}), and use less resources. In contrast, in this experiment, agents conditioned only to maximize task returns fail to correctly perform the task as they are unable to adapt to the teammate's characteristics. In addition, although the pure RL team unsurprisingly achieves the best performance, we note that this would require agents to know their exact teammates' profiles and co-training for each possible pair, while the ToMCAT agent can use a single ToMnet and MADiff model and adapt to a multitude of teammate behaviors. 

\subsubsection{Unknown Teammates}%

\begin{table}[!t]
    \centering
    \caption{ToM impact in the absence of teammate priors.}
    \label{Table:ExpUnknown}
    \begin{tabular}{l r @{\hspace{2pt}$\pm$\hspace{2pt}} r r @{\hspace{2pt}$\pm$\hspace{2pt}} r r @{\hspace{2pt}$\pm$\hspace{2pt}} r}
        \toprule
        \multicolumn{1}{l}{\textbf{Condition}}
        & \multicolumn{2}{c}{\textbf{Plan Count}}
        & \multicolumn{2}{c}{\textbf{Task Rwd}}
        & \multicolumn{2}{c}{\textbf{Indiv. Rwd}} \\
        \midrule

        \emph{Unconditioned} & $\boldsymbol{79.79}$ & $\boldsymbol{3.61}$ & $10.64$ & $1.83$ & $-34.32$ & $9.72$ \\
        \emph{Returns} & $105.11$ & $3.09$ & $10.88$ & $1.50$ & $-41.37$ & $10.58$ \\
        \emph{Profile} & $87.00$ & $3.95$ & $\boldsymbol{14.24}$ & $\boldsymbol{2.33}$ & $-31.29$ & $9.79$ \\
        \emph{ToM} & $83.97$ & $3.81$ & $12.52$ & $2.19$ & $-29.25$ & $9.34$ \\
        \emph{RL Unknown} & \multicolumn{2}{c}{-} & $10.36$ & $1.88$ & $\boldsymbol{-11.05}$ & $\boldsymbol{5.76}$ \\

        \bottomrule
    \end{tabular}
\end{table}


Here we want to address the question: \textit{``Can ToMCAT agents learn from and adapt to unknown teammates?''} We adopted a methodology similar to that of the previous experiment, but where the \textit{ToM} agent was \emph{not} provided any past trajectories about its teammate, \ie $\npast=0$. Instead, for the first $100$ timesteps of an episode, the agent updated a buffer of observations, and used that to compute the character embedding, $\echar$. We also used a different baseline against which to compare the MADiff agents, namely a team of RL agents that did not ``know'' their teammate. Recall from Sec.~\ref{Subsec:Data} that we co-trained RL policies via MARL for all pairs of agent profiles. Since we designed $7$ profiles, this results in $6$ policies trained for each profile. As such, for the \textit{RL Unknown} condition, we selected for each RL agent a policy from the corresponding set of profile policies uniformly at random.

Table~\ref{Table:ExpUnknown} shows the results for the different conditions. The pure RL agents failed to coordinate with their teammates due to the random policy selection. Notwithstanding, they attained the best mean individual performance since their policies were trained to maximize the reward function associated with their profiles, \emph{irrespectively} of the teammate. We also see that despite not being provided prior information about a teammate, the \textit{ToM} agent is still able to perform better than the unconditioned agent and the RL team (difference not statistically significant, $p\gg 0.05$). Moreover, the \textit{Profile} condition attained the highest mean task performance. 

The are a few possible reasons for the lower performance of the ToMCAT agent in this unknown teammate scenario. First, the MADiff models were trained with data produced by RL policies that have a predictable behavior when paired with the corresponding RL teammate---since their policies were co-trained---and were provided with $\npast=4$ past trajectories. However, here the Character embeddings are not informative enough, especially at the beginning of an episode. This results in the agents not coordinating their behavior, which in turn influences plan generation since we use history in-painting. This analysis is supported by Fig.~\ref{Fig:ResNoPriorProfProb}, where we see that when provided with prior information, a ToMCAT agent is able to correctly identify the teammate's profile ($P\approx1$) from the start. This contrasts with the \emph{No Prior} condition, where the probability increases as more observations are collected, but on average is no higher than $0.89$. We also tested updating the current observation buffer continuously beyond $t=100$ in a sliding window manner, but observed a decline in ToM performance, presumably because all past trajectories used to train the ToMnet and MADiff models were collected for the first $\pasttrajlen=100$ steps of episodes.

Overall, we would need to train the ToMnet and MADiff models from past trajectories collected at different timesteps, and vary the number of past trajectories during training, including providing no prior data, to increase the robustness of ToMCAT agents when paired with unknown teammates. In addition, we believe that training would benefit from data collected by pairing different agents, \eg randomly pairing RL agents, which would better capture the distribution of possible team behaviors in the task. We acknowledge the current limitations of our framework and leave these developments for future work.

\section{Conclusions \& Future Work}%
\label{Sec:Conclusions}

This paper presented ToMCAT, a new framework that combines ToM reasoning with multiagent planning for fast adaptation in complex multiagent tasks. It integrates a ToM network (ToMnet) that learns a strong prior of possible team behaviors and leverages meta-learning to make predictions about underlying characteristics of teammates from minimal data, with a multiagent denoising-diffusion policy (MADiff) approach that generates plans for the agent and its teammates conditioned by ToM embeddings of the team. An online dynamic replanning mechanism monitors plan execution and triggers the sampling of a new plan whenever there is a discrepancy between the plan and the current state of the world. We performed various experiments using different parameterizations of ToMCAT agents under different conditions in a simulated collaborative cooking domain. The results show the usefulness of the replanning mechanism in attaining good pairwise performance without the need to replan at short, fixed-time intervals. It also underscores the importance of reasoning about teammates' characteristics as informed by the ToM embeddings and the agent's own characteristics to allow for fast adaptation to different teammates, especially when no prior information is provided about them.

\textbf{Future work}: We are currently designing a joint model approach, where ToMnet and MADiff systems are combined into a probabilistic generative model that allows sampling multiagent plans conditioned on ToM information, while ``forcing'' the model to predict desired characteristics of teammate motivations and behaviors. We are also addressing the challenges of making ToMCAT agents more robust to unknown teammates and exploring how the framework can be applied in ad hoc teamwork settings. 
Finally, we plan to apply ToMCAT to learn from human data to infer profiles, behavior trends of people performing complex joint tasks, and explore its application in adversarial settings requiring nested ToM.





\bibliographystyle{ACM-Reference-Format} 
\bibliography{25_arxiv_tomcat.bib}


\begin{thebibliography}{31}


\ifx \showCODEN    \undefined \def \showCODEN     #1{\unskip}     \fi
\ifx \showDOI      \undefined \def \showDOI       #1{#1}\fi
\ifx \showISBNx    \undefined \def \showISBNx     #1{\unskip}     \fi
\ifx \showISBNxiii \undefined \def \showISBNxiii  #1{\unskip}     \fi
\ifx \showISSN     \undefined \def \showISSN      #1{\unskip}     \fi
\ifx \showLCCN     \undefined \def \showLCCN      #1{\unskip}     \fi
\ifx \shownote     \undefined \def \shownote      #1{#1}          \fi
\ifx \showarticletitle \undefined \def \showarticletitle #1{#1}   \fi
\ifx \showURL      \undefined \def \showURL       {\relax}        \fi
\providecommand\bibfield[2]{#2}
\providecommand\bibinfo[2]{#2}
\providecommand\natexlab[1]{#1}
\providecommand\showeprint[2][]{arXiv:#2}

\bibitem[\protect\citeauthoryear{Ajay, Du, Gupta, Tenenbaum, Jaakkola, and Agrawal}{Ajay et~al\mbox{.}}{2023}]%
        {ajay2023condgen}
\bibfield{author}{\bibinfo{person}{Anurag Ajay}, \bibinfo{person}{Yilun Du}, \bibinfo{person}{Abhi Gupta}, \bibinfo{person}{Joshua~B. Tenenbaum}, \bibinfo{person}{Tommi~S. Jaakkola}, {and} \bibinfo{person}{Pulkit Agrawal}.} \bibinfo{year}{2023}\natexlab{}.
\newblock \showarticletitle{Is Conditional Generative Modeling all you need for Decision Making?}. In \bibinfo{booktitle}{\emph{The Eleventh International Conference on Learning Representations}}.
\newblock
\urldef\tempurl%
\url{https://openreview.net/forum?id=sP1fo2K9DFG}
\showURL{%
\tempurl}


\bibitem[\protect\citeauthoryear{Baron-Cohen, Leslie, and Frith}{Baron-Cohen et~al\mbox{.}}{1985}]%
        {baroncohen1985tom}
\bibfield{author}{\bibinfo{person}{Simon Baron-Cohen}, \bibinfo{person}{Alan~M. Leslie}, {and} \bibinfo{person}{Uta Frith}.} \bibinfo{year}{1985}\natexlab{}.
\newblock \showarticletitle{Does the autistic child have a “theory of mind”?}
\newblock \bibinfo{journal}{\emph{Cognition}} \bibinfo{volume}{21}, \bibinfo{number}{1} (\bibinfo{year}{1985}), \bibinfo{pages}{37--46}.
\newblock
\showISSN{0010-0277}
\urldef\tempurl%
\url{https://doi.org/10.1016/0010-0277(85)90022-8}
\showDOI{\tempurl}


\bibitem[\protect\citeauthoryear{Carroll, Shah, Ho, Griffiths, Seshia, Abbeel, and Dragan}{Carroll et~al\mbox{.}}{2019}]%
        {carroll2019overcooked}
\bibfield{author}{\bibinfo{person}{Micah Carroll}, \bibinfo{person}{Rohin Shah}, \bibinfo{person}{Mark~K Ho}, \bibinfo{person}{Tom Griffiths}, \bibinfo{person}{Sanjit Seshia}, \bibinfo{person}{Pieter Abbeel}, {and} \bibinfo{person}{Anca Dragan}.} \bibinfo{year}{2019}\natexlab{}.
\newblock \showarticletitle{On the Utility of Learning about Humans for Human-AI Coordination}. In \bibinfo{booktitle}{\emph{Advances in Neural Information Processing Systems}}, \bibfield{editor}{\bibinfo{person}{H.~Wallach}, \bibinfo{person}{H.~Larochelle}, \bibinfo{person}{A.~Beygelzimer}, \bibinfo{person}{F.~d\textquotesingle Alch\'{e}-Buc}, \bibinfo{person}{E.~Fox}, {and} \bibinfo{person}{R.~Garnett}} (Eds.), Vol.~\bibinfo{volume}{32}. \bibinfo{publisher}{Curran Associates, Inc.}
\newblock
\urldef\tempurl%
\url{https://proceedings.neurips.cc/paper_files/paper/2019/file/f5b1b89d98b7286673128a5fb112cb9a-Paper.pdf}
\showURL{%
\tempurl}


\bibitem[\protect\citeauthoryear{Dafoe, Hughes, Bachrach, Collins, McKee, Leibo, Larson, and Graepel}{Dafoe et~al\mbox{.}}{2020}]%
        {dafoe2020coopai}
\bibfield{author}{\bibinfo{person}{Allan Dafoe}, \bibinfo{person}{Edward Hughes}, \bibinfo{person}{Yoram Bachrach}, \bibinfo{person}{Tantum Collins}, \bibinfo{person}{Kevin~R. McKee}, \bibinfo{person}{Joel~Z. Leibo}, \bibinfo{person}{Kate Larson}, {and} \bibinfo{person}{Thore Graepel}.} \bibinfo{year}{2020}\natexlab{}.
\newblock \bibinfo{title}{Open Problems in Cooperative AI}.
\newblock
\newblock
\showeprint[arxiv]{2012.08630}~[cs.AI]
\urldef\tempurl%
\url{https://arxiv.org/abs/2012.08630}
\showURL{%
\tempurl}


\bibitem[\protect\citeauthoryear{Gmytrasiewicz and Doshi}{Gmytrasiewicz and Doshi}{2005}]%
        {gmytrasiewicz2005ipomdp}
\bibfield{author}{\bibinfo{person}{Piotr~J Gmytrasiewicz} {and} \bibinfo{person}{Prashant Doshi}.} \bibinfo{year}{2005}\natexlab{}.
\newblock \showarticletitle{{A framework for sequential planning in multi-agent settings}}.
\newblock \bibinfo{journal}{\emph{J. Artif. Int. Res.}} \bibinfo{volume}{24}, \bibinfo{number}{1} (\bibinfo{date}{jul} \bibinfo{year}{2005}), \bibinfo{pages}{49--79}.
\newblock
\showISSN{1076-9757}


\bibitem[\protect\citeauthoryear{Gu, Zhao, Hao, and An}{Gu et~al\mbox{.}}{2022}]%
        {gu2022online}
\bibfield{author}{\bibinfo{person}{Pengjie Gu}, \bibinfo{person}{Mengchen Zhao}, \bibinfo{person}{Jianye Hao}, {and} \bibinfo{person}{Bo An}.} \bibinfo{year}{2022}\natexlab{}.
\newblock \showarticletitle{Online Ad Hoc Teamwork under Partial Observability}. In \bibinfo{booktitle}{\emph{International Conference on Learning Representations}}.
\newblock
\urldef\tempurl%
\url{https://openreview.net/forum?id=18Ys0-PzyPI}
\showURL{%
\tempurl}


\bibitem[\protect\citeauthoryear{Gurney, Marsella, Ustun, and Pynadath}{Gurney et~al\mbox{.}}{2022}]%
        {gurney2022tom}
\bibfield{author}{\bibinfo{person}{Nikolos Gurney}, \bibinfo{person}{Stacy Marsella}, \bibinfo{person}{Volkan Ustun}, {and} \bibinfo{person}{David~V. Pynadath}.} \bibinfo{year}{2022}\natexlab{}.
\newblock \showarticletitle{Operationalizing Theories of Theory of Mind: A Survey}. In \bibinfo{booktitle}{\emph{Computational Theory of Mind for Human-Machine Teams}} \emph{(\bibinfo{series}{AAAI-FSS 2021})}, \bibfield{editor}{\bibinfo{person}{Nikolos Gurney} {and} \bibinfo{person}{Gita Sukthankar}} (Eds.). \bibinfo{publisher}{Springer Nature Switzerland}, \bibinfo{address}{Cham}, \bibinfo{pages}{3--20}.
\newblock
\showISBNx{978-3-031-21671-8}
\urldef\tempurl%
\url{https://doi.org/10.1007/978-3-031-21671-8_1}
\showDOI{\tempurl}


\bibitem[\protect\citeauthoryear{Han and Gmytrasiewicz}{Han and Gmytrasiewicz}{2018}]%
        {han2018learning}
\bibfield{author}{\bibinfo{person}{Yanlin Han} {and} \bibinfo{person}{Piotr Gmytrasiewicz}.} \bibinfo{year}{2018}\natexlab{}.
\newblock \showarticletitle{{Learning others' intentional models in multi-agent settings using interactive POMDPs}}.
\newblock \bibinfo{journal}{\emph{Advances in Neural Information Processing Systems}}  \bibinfo{volume}{31} (\bibinfo{year}{2018}).
\newblock


\bibitem[\protect\citeauthoryear{Han and Gmytrasiewicz}{Han and Gmytrasiewicz}{2019}]%
        {han2019ipomdp}
\bibfield{author}{\bibinfo{person}{Yanlin Han} {and} \bibinfo{person}{Piotr Gmytrasiewicz}.} \bibinfo{year}{2019}\natexlab{}.
\newblock \showarticletitle{{Ipomdp-net: A deep neural network for partially observable multi-agent planning using interactive pomdps}}. In \bibinfo{booktitle}{\emph{Proceedings of the AAAI Conference on Artificial Intelligence}}, Vol.~\bibinfo{volume}{33}. \bibinfo{pages}{6062--6069}.
\newblock


\bibitem[\protect\citeauthoryear{Harsanyi}{Harsanyi}{1967}]%
        {harsanyi1967bayes}
\bibfield{author}{\bibinfo{person}{John~C. Harsanyi}.} \bibinfo{year}{1967}\natexlab{}.
\newblock \showarticletitle{{Games with Incomplete Information Played by “Bayesian” Players, I–III Part I. The Basic Model}}.
\newblock \bibinfo{journal}{\emph{Management Science}} \bibinfo{volume}{14}, \bibinfo{number}{3} (\bibinfo{date}{nov} \bibinfo{year}{1967}), \bibinfo{pages}{159--182}.
\newblock
\showISSN{0025-1909}
\urldef\tempurl%
\url{https://doi.org/10.1287/mnsc.14.3.159}
\showDOI{\tempurl}


\bibitem[\protect\citeauthoryear{Ho, Jain, and Abbeel}{Ho et~al\mbox{.}}{2020}]%
        {ho2020diff}
\bibfield{author}{\bibinfo{person}{Jonathan Ho}, \bibinfo{person}{Ajay Jain}, {and} \bibinfo{person}{Pieter Abbeel}.} \bibinfo{year}{2020}\natexlab{}.
\newblock \showarticletitle{Denoising Diffusion Probabilistic Models}. In \bibinfo{booktitle}{\emph{Advances in Neural Information Processing Systems}}, \bibfield{editor}{\bibinfo{person}{H.~Larochelle}, \bibinfo{person}{M.~Ranzato}, \bibinfo{person}{R.~Hadsell}, \bibinfo{person}{M.F. Balcan}, {and} \bibinfo{person}{H.~Lin}} (Eds.), Vol.~\bibinfo{volume}{33}. \bibinfo{publisher}{Curran Associates, Inc.}, \bibinfo{pages}{6840--6851}.
\newblock
\urldef\tempurl%
\url{https://proceedings.neurips.cc/paper_files/paper/2020/file/4c5bcfec8584af0d967f1ab10179ca4b-Paper.pdf}
\showURL{%
\tempurl}


\bibitem[\protect\citeauthoryear{Ho and Salimans}{Ho and Salimans}{2022}]%
        {ho2022guidance}
\bibfield{author}{\bibinfo{person}{Jonathan Ho} {and} \bibinfo{person}{Tim Salimans}.} \bibinfo{year}{2022}\natexlab{}.
\newblock \bibinfo{title}{Classifier-Free Diffusion Guidance}.
\newblock
\newblock
\showeprint[arxiv]{2207.12598}~[cs.LG]
\urldef\tempurl%
\url{https://arxiv.org/abs/2207.12598}
\showURL{%
\tempurl}


\bibitem[\protect\citeauthoryear{Jaderberg, Czarnecki, Dunning, Marris, Lever, Castaneda, Beattie, Rabinowitz, Morcos, Ruderman, et~al\mbox{.}}{Jaderberg et~al\mbox{.}}{2019}]%
        {jaderberg2019human}
\bibfield{author}{\bibinfo{person}{Max Jaderberg}, \bibinfo{person}{Wojciech~M Czarnecki}, \bibinfo{person}{Iain Dunning}, \bibinfo{person}{Luke Marris}, \bibinfo{person}{Guy Lever}, \bibinfo{person}{Antonio~Garcia Castaneda}, \bibinfo{person}{Charles Beattie}, \bibinfo{person}{Neil~C Rabinowitz}, \bibinfo{person}{Ari~S Morcos}, \bibinfo{person}{Avraham Ruderman}, {et~al\mbox{.}}} \bibinfo{year}{2019}\natexlab{}.
\newblock \showarticletitle{Human-level performance in 3D multiplayer games with population-based reinforcement learning}.
\newblock \bibinfo{journal}{\emph{Science}} \bibinfo{volume}{364}, \bibinfo{number}{6443} (\bibinfo{year}{2019}), \bibinfo{pages}{859--865}.
\newblock


\bibitem[\protect\citeauthoryear{Janner, Du, Tenenbaum, and Levine}{Janner et~al\mbox{.}}{2022}]%
        {janner2022diff}
\bibfield{author}{\bibinfo{person}{Michael Janner}, \bibinfo{person}{Yilun Du}, \bibinfo{person}{Joshua Tenenbaum}, {and} \bibinfo{person}{Sergey Levine}.} \bibinfo{year}{2022}\natexlab{}.
\newblock \showarticletitle{Planning with Diffusion for Flexible Behavior Synthesis}. In \bibinfo{booktitle}{\emph{Proceedings of the 39th International Conference on Machine Learning}} \emph{(\bibinfo{series}{Proceedings of Machine Learning Research}, Vol.~\bibinfo{volume}{162})}, \bibfield{editor}{\bibinfo{person}{Kamalika Chaudhuri}, \bibinfo{person}{Stefanie Jegelka}, \bibinfo{person}{Le~Song}, \bibinfo{person}{Csaba Szepesvari}, \bibinfo{person}{Gang Niu}, {and} \bibinfo{person}{Sivan Sabato}} (Eds.). \bibinfo{publisher}{PMLR}, \bibinfo{pages}{9902--9915}.
\newblock
\urldef\tempurl%
\url{https://proceedings.mlr.press/v162/janner22a.html}
\showURL{%
\tempurl}


\bibitem[\protect\citeauthoryear{Jara-Ettinger}{Jara-Ettinger}{2019}]%
        {jaraettinger2019tomirl}
\bibfield{author}{\bibinfo{person}{Julian Jara-Ettinger}.} \bibinfo{year}{2019}\natexlab{}.
\newblock \showarticletitle{Theory of mind as inverse reinforcement learning}.
\newblock \bibinfo{journal}{\emph{Current Opinion in Behavioral Sciences}}  \bibinfo{volume}{29} (\bibinfo{year}{2019}), \bibinfo{pages}{105--110}.
\newblock
\showISSN{2352-1546}
\urldef\tempurl%
\url{https://doi.org/10.1016/j.cobeha.2019.04.010}
\showDOI{\tempurl}
\newblock
\shownote{Artificial Intelligence.}


\bibitem[\protect\citeauthoryear{Karkus, Hsu, and Lee}{Karkus et~al\mbox{.}}{2017}]%
        {Karkus2017qmdpnet}
\bibfield{author}{\bibinfo{person}{Peter Karkus}, \bibinfo{person}{David Hsu}, {and} \bibinfo{person}{Wee~Sun Lee}.} \bibinfo{year}{2017}\natexlab{}.
\newblock \showarticletitle{{QMDP-Net: Deep Learning for Planning under Partial Observability}}. In \bibinfo{booktitle}{\emph{Proceedings of the 31st International Conference on Neural Information Processing Systems}} \emph{(\bibinfo{series}{NIPS'17})}. \bibinfo{address}{Long Beach, California, USA}, \bibinfo{pages}{4697--4707}.
\newblock
\urldef\tempurl%
\url{https://doi.org/10.5555/3295222.3295223}
\showDOI{\tempurl}


\bibitem[\protect\citeauthoryear{Lee, {Ganapathi Subramanian}, and Crowley}{Lee et~al\mbox{.}}{2022}]%
        {lee2022investigation}
\bibfield{author}{\bibinfo{person}{Ken~Ming Lee}, \bibinfo{person}{Sriram {Ganapathi Subramanian}}, {and} \bibinfo{person}{Mark Crowley}.} \bibinfo{year}{2022}\natexlab{}.
\newblock \showarticletitle{{Investigation of independent reinforcement learning algorithms in multi-agent environments}}.
\newblock \bibinfo{journal}{\emph{Frontiers in Artificial Intelligence}}  \bibinfo{volume}{5} (\bibinfo{year}{2022}), \bibinfo{pages}{805823}.
\newblock


\bibitem[\protect\citeauthoryear{Lowe, Wu, Tamar, Harb, Abbeel, and Mordatch}{Lowe et~al\mbox{.}}{2017}]%
        {lowe2017maac}
\bibfield{author}{\bibinfo{person}{Ryan Lowe}, \bibinfo{person}{Yi Wu}, \bibinfo{person}{Aviv Tamar}, \bibinfo{person}{Jean Harb}, \bibinfo{person}{Pieter Abbeel}, {and} \bibinfo{person}{Igor Mordatch}.} \bibinfo{year}{2017}\natexlab{}.
\newblock \showarticletitle{{Multi-Agent Actor-Critic for Mixed Cooperative-Competitive Environments}}. In \bibinfo{booktitle}{\emph{Proceedings of the 31st International Conference on Neural Information Processing Systems}} \emph{(\bibinfo{series}{NIPS'17})}. \bibinfo{publisher}{Curran Associates Inc.}, \bibinfo{address}{Red Hook, NY, USA}, \bibinfo{pages}{6382--6393}.
\newblock
\showISBNx{9781510860964}


\bibitem[\protect\citeauthoryear{Pynadath and Marsella}{Pynadath and Marsella}{2005}]%
        {pynadath2005psychsim}
\bibfield{author}{\bibinfo{person}{David~V. Pynadath} {and} \bibinfo{person}{Stacy~C. Marsella}.} \bibinfo{year}{2005}\natexlab{}.
\newblock \showarticletitle{PsychSim: modeling theory of mind with decision-theoretic agents}. In \bibinfo{booktitle}{\emph{Proceedings of the 19th International Joint Conference on Artificial Intelligence}} (Edinburgh, Scotland) \emph{(\bibinfo{series}{IJCAI'05})}. \bibinfo{publisher}{Morgan Kaufmann Publishers Inc.}, \bibinfo{address}{San Francisco, CA, USA}, \bibinfo{pages}{1181–1186}.
\newblock


\bibitem[\protect\citeauthoryear{Rabinowitz, Perbet, Song, Zhang, Eslami, and Botvinick}{Rabinowitz et~al\mbox{.}}{2018}]%
        {rabinowitz2018tomnet}
\bibfield{author}{\bibinfo{person}{Neil Rabinowitz}, \bibinfo{person}{Frank Perbet}, \bibinfo{person}{Francis Song}, \bibinfo{person}{Chiyuan Zhang}, \bibinfo{person}{S.~M.~Ali Eslami}, {and} \bibinfo{person}{Matthew Botvinick}.} \bibinfo{year}{2018}\natexlab{}.
\newblock \showarticletitle{Machine Theory of Mind}. In \bibinfo{booktitle}{\emph{Proceedings of the 35th International Conference on Machine Learning}} \emph{(\bibinfo{series}{Proceedings of Machine Learning Research}, Vol.~\bibinfo{volume}{80})}, \bibfield{editor}{\bibinfo{person}{Jennifer Dy} {and} \bibinfo{person}{Andreas Krause}} (Eds.). \bibinfo{publisher}{PMLR}, \bibinfo{pages}{4218--4227}.
\newblock
\urldef\tempurl%
\url{https://proceedings.mlr.press/v80/rabinowitz18a.html}
\showURL{%
\tempurl}


\bibitem[\protect\citeauthoryear{Ribeiro, Martinho, Sardinha, and Melo}{Ribeiro et~al\mbox{.}}{2023}]%
        {ribeiro2023adhoc}
\bibfield{author}{\bibinfo{person}{Jo{\~{a}}o~G. Ribeiro}, \bibinfo{person}{Cassandro Martinho}, \bibinfo{person}{Alberto Sardinha}, {and} \bibinfo{person}{Francisco~S. Melo}.} \bibinfo{year}{2023}\natexlab{}.
\newblock \showarticletitle{{Making Friends in the Dark: Ad Hoc Teamwork Under Partial Observability}}.
\newblock \bibinfo{journal}{\emph{Frontiers in Artificial Intelligence and Applications}}  \bibinfo{volume}{372} (\bibinfo{date}{sep} \bibinfo{year}{2023}), \bibinfo{pages}{1954--1961}.
\newblock
\showISBNx{9781643684369}
\showISSN{18798314}
\urldef\tempurl%
\url{https://doi.org/10.3233/FAIA230486}
\showDOI{\tempurl}


\bibitem[\protect\citeauthoryear{Rocha, da~Silva, Morales, Sarkadi, and Panisson}{Rocha et~al\mbox{.}}{2023}]%
        {rocha2023tom}
\bibfield{author}{\bibinfo{person}{Michele Rocha}, \bibinfo{person}{Heitor~Henrique da Silva}, \bibinfo{person}{Anal{\'u}cia~Schiaffino Morales}, \bibinfo{person}{Stefan Sarkadi}, {and} \bibinfo{person}{Alison~R. Panisson}.} \bibinfo{year}{2023}\natexlab{}.
\newblock \showarticletitle{Applying Theory of Mind to Multi-agent Systems: A Systematic Review}. In \bibinfo{booktitle}{\emph{Intelligent Systems}} \emph{(\bibinfo{series}{BRACIS 2023})}, \bibfield{editor}{\bibinfo{person}{Murilo~C. Naldi} {and} \bibinfo{person}{Reinaldo A.~C. Bianchi}} (Eds.). \bibinfo{publisher}{Springer Nature Switzerland}, \bibinfo{address}{Cham}, \bibinfo{pages}{367--381}.
\newblock
\showISBNx{978-3-031-45368-7}
\urldef\tempurl%
\url{https://doi.org/10.1007/978-3-031-45368-7_24}
\showDOI{\tempurl}


\bibitem[\protect\citeauthoryear{Schulman, Wolski, Dhariwal, Radford, and Klimov}{Schulman et~al\mbox{.}}{2017}]%
        {schulman2017ppo}
\bibfield{author}{\bibinfo{person}{John Schulman}, \bibinfo{person}{Filip Wolski}, \bibinfo{person}{Prafulla Dhariwal}, \bibinfo{person}{Alec Radford}, {and} \bibinfo{person}{Oleg Klimov}.} \bibinfo{year}{2017}\natexlab{}.
\newblock \bibinfo{title}{Proximal Policy Optimization Algorithms}.
\newblock
\newblock
\showeprint[arxiv]{1707.06347}~[cs.LG]
\urldef\tempurl%
\url{https://arxiv.org/abs/1707.06347}
\showURL{%
\tempurl}


\bibitem[\protect\citeauthoryear{Silver, Hubert, Schrittwieser, Antonoglou, Lai, Guez, Lanctot, Sifre, Kumaran, Graepel, et~al\mbox{.}}{Silver et~al\mbox{.}}{2017}]%
        {silver2017mastering}
\bibfield{author}{\bibinfo{person}{David Silver}, \bibinfo{person}{Thomas Hubert}, \bibinfo{person}{Julian Schrittwieser}, \bibinfo{person}{Ioannis Antonoglou}, \bibinfo{person}{Matthew Lai}, \bibinfo{person}{Arthur Guez}, \bibinfo{person}{Marc Lanctot}, \bibinfo{person}{Laurent Sifre}, \bibinfo{person}{Dharshan Kumaran}, \bibinfo{person}{Thore Graepel}, {et~al\mbox{.}}} \bibinfo{year}{2017}\natexlab{}.
\newblock \showarticletitle{Mastering chess and shogi by self-play with a general reinforcement learning algorithm}.
\newblock \bibinfo{journal}{\emph{arXiv preprint arXiv:1712.01815}} (\bibinfo{year}{2017}).
\newblock


\bibitem[\protect\citeauthoryear{Sohl-Dickstein, Weiss, Maheswaranathan, and Ganguli}{Sohl-Dickstein et~al\mbox{.}}{2015}]%
        {sohldickstein2015diff}
\bibfield{author}{\bibinfo{person}{Jascha Sohl-Dickstein}, \bibinfo{person}{Eric Weiss}, \bibinfo{person}{Niru Maheswaranathan}, {and} \bibinfo{person}{Surya Ganguli}.} \bibinfo{year}{2015}\natexlab{}.
\newblock \showarticletitle{Deep Unsupervised Learning using Nonequilibrium Thermodynamics}. In \bibinfo{booktitle}{\emph{Proceedings of the 32nd International Conference on Machine Learning}} \emph{(\bibinfo{series}{Proceedings of Machine Learning Research}, Vol.~\bibinfo{volume}{37})}, \bibfield{editor}{\bibinfo{person}{Francis Bach} {and} \bibinfo{person}{David Blei}} (Eds.). \bibinfo{publisher}{PMLR}, \bibinfo{address}{Lille, France}, \bibinfo{pages}{2256--2265}.
\newblock
\urldef\tempurl%
\url{https://proceedings.mlr.press/v37/sohl-dickstein15.html}
\showURL{%
\tempurl}


\bibitem[\protect\citeauthoryear{Song, Meng, and Ermon}{Song et~al\mbox{.}}{2021}]%
        {song2021ddim}
\bibfield{author}{\bibinfo{person}{Jiaming Song}, \bibinfo{person}{Chenlin Meng}, {and} \bibinfo{person}{Stefano Ermon}.} \bibinfo{year}{2021}\natexlab{}.
\newblock \showarticletitle{Denoising Diffusion Implicit Models}. In \bibinfo{booktitle}{\emph{International Conference on Learning Representations}}.
\newblock
\urldef\tempurl%
\url{https://arxiv.org/abs/2010.02502}
\showURL{%
\tempurl}


\bibitem[\protect\citeauthoryear{Stone, Kaminka, Kraus, and Rosenschein}{Stone et~al\mbox{.}}{2010}]%
        {stone2010adhoc}
\bibfield{author}{\bibinfo{person}{Peter Stone}, \bibinfo{person}{Gal Kaminka}, \bibinfo{person}{Sarit Kraus}, {and} \bibinfo{person}{Jeffrey Rosenschein}.} \bibinfo{year}{2010}\natexlab{}.
\newblock \showarticletitle{{Ad Hoc Autonomous Agent Teams: Collaboration without Pre-Coordination}}.
\newblock \bibinfo{journal}{\emph{Proceedings of the AAAI Conference on Artificial Intelligence}} \bibinfo{volume}{24}, \bibinfo{number}{1} (\bibinfo{date}{jul} \bibinfo{year}{2010}), \bibinfo{pages}{1504--1509}.
\newblock
\showISSN{2374-3468}
\urldef\tempurl%
\url{https://doi.org/10.1609/aaai.v24i1.7529}
\showDOI{\tempurl}


\bibitem[\protect\citeauthoryear{Sutton, McAllester, Singh, and Mansour}{Sutton et~al\mbox{.}}{1999}]%
        {sutton1999functionapprox}
\bibfield{author}{\bibinfo{person}{Richard~S Sutton}, \bibinfo{person}{David McAllester}, \bibinfo{person}{Satinder Singh}, {and} \bibinfo{person}{Yishay Mansour}.} \bibinfo{year}{1999}\natexlab{}.
\newblock \showarticletitle{{Policy Gradient Methods for Reinforcement Learning with Function Approximation}}. In \bibinfo{booktitle}{\emph{Proceedings of the 12th International Conference on Neural Information Processing Systems}} \emph{(\bibinfo{series}{NIPS'99})}. \bibinfo{publisher}{MIT Press}, \bibinfo{address}{Cambridge, MA, USA}, \bibinfo{pages}{1057--1063}.
\newblock


\bibitem[\protect\citeauthoryear{Wu, Sequeira, and Pynadath}{Wu et~al\mbox{.}}{2023}]%
        {wu2023multiagent}
\bibfield{author}{\bibinfo{person}{Haochen Wu}, \bibinfo{person}{Pedro Sequeira}, {and} \bibinfo{person}{David~V Pynadath}.} \bibinfo{year}{2023}\natexlab{}.
\newblock \showarticletitle{{Multiagent Inverse Reinforcement Learning via Theory of Mind Reasoning}}. In \bibinfo{booktitle}{\emph{Proceedings of the 2023 International Conference on Autonomous Agents and Multiagent Systems}}. \bibinfo{pages}{708--716}.
\newblock


\bibitem[\protect\citeauthoryear{Wu, Wang, Evans, Tenenbaum, Parkes, and Kleiman-Weiner}{Wu et~al\mbox{.}}{2021}]%
        {wu2021cooks}
\bibfield{author}{\bibinfo{person}{Sarah~A. Wu}, \bibinfo{person}{Rose~E. Wang}, \bibinfo{person}{James~A. Evans}, \bibinfo{person}{Joshua~B. Tenenbaum}, \bibinfo{person}{David~C. Parkes}, {and} \bibinfo{person}{Max Kleiman-Weiner}.} \bibinfo{year}{2021}\natexlab{}.
\newblock \showarticletitle{Too Many Cooks: Bayesian Inference for Coordinating Multi-Agent Collaboration}.
\newblock \bibinfo{journal}{\emph{Topics in Cognitive Science}} \bibinfo{volume}{13}, \bibinfo{number}{2} (\bibinfo{year}{2021}), \bibinfo{pages}{414--432}.
\newblock
\urldef\tempurl%
\url{https://doi.org/10.1111/tops.12525}
\showDOI{\tempurl}
\showeprint{https://onlinelibrary.wiley.com/doi/pdf/10.1111/tops.12525}


\bibitem[\protect\citeauthoryear{Zhu, Liu, Mao, Kang, Xu, Yu, Ermon, and Zhang}{Zhu et~al\mbox{.}}{2024}]%
        {zhu2024madiff}
\bibfield{author}{\bibinfo{person}{Zhengbang Zhu}, \bibinfo{person}{Minghuan Liu}, \bibinfo{person}{Liyuan Mao}, \bibinfo{person}{Bingyi Kang}, \bibinfo{person}{Minkai Xu}, \bibinfo{person}{Yong Yu}, \bibinfo{person}{Stefano Ermon}, {and} \bibinfo{person}{Weinan Zhang}.} \bibinfo{year}{2024}\natexlab{}.
\newblock \showarticletitle{{MAD}iff: Offline Multi-agent Learning with Diffusion Models}. In \bibinfo{booktitle}{\emph{The Thirty-eighth Annual Conference on Neural Information Processing Systems}}.
\newblock
\urldef\tempurl%
\url{https://openreview.net/forum?id=PvoxbjcRPT}
\showURL{%
\tempurl}


\end{thebibliography}


\clearpage
\appendix
\renewcommand{\thetable}{A.\arabic{table}}
\setcounter{table}{0}
\renewcommand{\thefigure}{A.\arabic{figure}}
\setcounter{figure}{0}

\section{Appendix}%
\label{Sec:Appendix}

Here we provide additional implementation details and results.

\subsection{Dynamic Replanning Algorithm}%
\label{Subsec:AppDynReplan}

Our dynamic replanning algorithm (Alg.~\ref{Alg:Planning}) unfolds as follows: for a given agent $i$, we first initialize the character embeddings for teammates from past trajectories, if available. We also initialize empty queues for the observation history, $h$, and the planned sequence of observations, $\trajplan$. 

At each step, after the agent makes a new observation in the environment, we check if replanning is needed (line 4). If a plan has been previously generated and is not depleted, we compute the mean squared error between the current observation and the observation predicted in the plan, \ie we set $\obsdiff \coloneq \norm{\obsnorm(\obs_i) - \obsnorm(\hat{\obs_i})}^2$, where $\obsnorm:\observations \to [0,1]$ is an observation normalizer detailed in Sec.~\ref{Subsec:AppMADiffTrain}. 

If the observation difference is greater than $\obsthresh=0.2$, we generate a new plan (lines 5--11). First, the mental embedding is computed from the current trajectory, and all conditioning variables---any combination of returns ($\rwdfunc_i(\traj{i})$), observer profile ($\profile_i$), character embedding ($\echar$), mental embedding ($\emental$)---are concatenated into $\conditions{\traj{i}}$ (line 5). We then generate a new plan through conditional sampling, starting with $\jointobstraj[\diffsteps] \sim \normal{\boldsymbol{0}}{\boldsymbol{I}}$. At each diffusion step $k$, the portion of the observer agent in $\jointobstraj[k]$ is ``in-painted'' with the corresponding current observation and $\history$ steps of observation history (line 8). We iteratively refine (denoise) trajectories $\jointobstraj[k]$ using the perturbed noise $\hat{\noise}$ following \cite{zhu2024madiff,ajay2023condgen}, where $\noisemodel(\jointobstraj[k], k)$ is the unconditional noise, $\condguide$ is a scalar for extracting the distinct portions of the trajectory that exhibit $\conditions{\traj{i}}$, and $\texttt{denoise}$ is the noise scheduler predicting the previous sample, $\jointobstraj[k-1]$, from $\jointobstraj[k]$ by propagating the diffusion process from $\hat{\noise}$. 

Once the final, denoised trajectory, $\jointobstraj[0]$, is generated, we discard the first $\history+1$ steps---since they were only used to condition the generation process---and extract agent $i$'s portion, $\hat{\traj{}}_i$, re-initializing the plan queue, $\trajplan$ (line 11). The agent then retrieves an action by using the inverse dynamics model $\invdyn$ on consecutive observations extracted from the plan (line 12).%
\footnote{Here the $\texttt{dequeue}$ function extracts observations from the queue in a FIFO manner.}

\subsection{Agent Profiles}%
\label{Subsec:AppProfiles}

\begin{table*}[!ht]
    \centering
    \caption{The different reward profiles used in our experiments.}
    \label{Table:Profiles}
    \begin{tabular}{l R{35pt} R{35pt} R{35pt} R{35pt} R{35pt} R{35pt} R{35pt}}
        \toprule
        \multicolumn{1}{l}{\textbf{Feature}, $\boldsymbol{\features_k}$}
        & \multicolumn{1}{r}{\textbf{Cook}}
        & \multicolumn{1}{r}{\textbf{Server}}
        & \multicolumn{1}{r}{\textbf{Helper}}
        & \multicolumn{1}{r}{\textbf{Far Helper}}
        & \multicolumn{1}{r}{\textbf{Follower}}
        & \multicolumn{1}{r}{\textbf{Sparse}}
        & \multicolumn{1}{r}{\textbf{Random}} \\
        \midrule

        \emph{Onion Drop} & $-5.00$ & $1.00$ & $0.00$ & $0.00$ & $0.00$ & $0.00$ & $0.00$ \\
        \emph{Onion Pickup} & $1.00$ & $-5.00$ & $0.10$ & $0.10$ & $0.00$ & $0.00$ & $0.00$ \\
        \emph{Dish Drop} & $1.00$ & $-5.00$ & $0.00$ & $0.00$ & $0.00$ & $0.00$ & $0.00$ \\
        \emph{Dish Pickup} & $-5.00$ & $1.00$ & $0.10$ & $0.10$ & $0.00$ & $0.00$ & $0.00$ \\
        \emph{Potting Onion} & $5.00$ & $-1.00$ & $10.00$ & $10.00$ & $0.00$ & $0.00$ & $0.00$ \\
        \emph{Soup Delivery} & $0.00$ & $5.00$ & $0.00$ & $0.00$ & $0.00$ & $0.00$ & $0.00$ \\
        \emph{Soup Drop} & $1.00$ & $-5.00$ & $-20.00$ & $-20.00$ & $0.00$ & $0.00$ & $0.00$ \\
        \emph{Soup Pickup} & $-5.00$ & $10.00$ & $15.00$ & $15.00$ & $0.00$ & $0.00$ & $0.00$ \\
        \emph{Self Pos X} & $0.00$ & $0.00$ & $0.00$ & $0.00$ & $0.00$ & $0.00$ & $0.00$ \\
        \emph{Self Pos Y} & $0.00$ & $0.00$ & $0.00$ & $0.00$ & $0.00$ & $0.00$ & $0.00$ \\
        \emph{Dist To Other Player X} & $0.00$ & $0.00$ & $0.00$ & $0.00$ & $0.00$ & $0.00$ & $0.00$ \\
        \emph{Dist To Other Player Y} & $0.00$ & $0.00$ & $0.00$ & $0.00$ & $0.00$ & $0.00$ & $0.00$ \\
        \emph{Path Dist To Other Player} & $0.00$ & $0.00$ & $0.00$ & $0.01$ & $-1.00$ & $0.00$ & $0.00$ \\
        \emph{Task Rew Frac} & $1.00$ & $1.00$ & $1.00$ & $1.00$ & $0.00$ & $1.00$ & $0.00$ \\

        \bottomrule
    \end{tabular}
\end{table*}

The profiles used in our experiments are listed in Table~\ref{Table:Profiles}, where for each profile (column) we list the weight $\rwdweights_k$ (row value) associated with each feature $\features_k$. \emph{Drop}, \emph{Pickup}, and \emph{Delivery} correspond to binary features indicating whether the agent's action at the previous timestep triggered the corresponding event, while \emph{Potting Onion} assesses whether an onion was put in a pot by the agent (again corresponding to agent's previous action). \emph{X} and \emph{Y} are numeric features corresponding to agent locations and relative distances to the teammate in the gridworld, and \emph{Path Dist} is the Manhattan distance to the teammate. \emph{Task Rew Frac} is the fraction of the task reward received by the agent, corresponding to a reward of $\rwdtask=20$ if \emph{any} agent successfully delivers a soup at timestep $t$, and $0$ otherwise.

\subsection{Multiagent RL Training}%
\label{Subsec:AppMARL}

For training, we adopted the PPO algorithm \citep{schulman2017ppo} implementation in Ray's RLLib toolkit.%
\footnote{\url{https://docs.ray.io/en/latest/rllib/index.html}}
Each pair was trained for $1\,000$ iterations with $64$ parallel workers. We used a learning rate of $8\times10^{-4}$ and a linearly decreasing entropy loss coefficient. At each iteration, a batch of data was collected from workers, each spwaning the overcooked environment with the layout shown in Fig.~\ref{Fig:Overcooked}, randomizing the initial locations of agents. Each episode ran for a period of $400$ timesteps. We collected $25\,600$ timesteps per batch, with a mini batch size of $6\,400$ for gradient updates, training for $420$ iterations of $8$ epochs each. 

\subsection{ToMnet}%
\label{Subsec:AppToMNET}

\subsubsection{Dataset/Feature Preparation}

We started with the pairwise RL agent dataset, $\trajset$, generated as described above. We sampled $2\,000$ datapoints for each agent/profile to train the ToMnet, where in each trial we selected a teammate uniformly at random. Each datapoint consisted of the following: we randomly sampled $\npast=4$ trajectories of the agent pair of $\pasttrajlen=100$ steps from $\trajset$ to serve as ``past'' trajectories for the Character net; we sampled an additional trajectory and split it at a timestep, $t \sim \uniform{1}{\trajlen-30}$, selected uniformly at random. The observation of the observer agent at $t$ was used as the ``current'' observation to be fed to the Prediction net, while the past $\curtrajlen = 10$ timesteps acted as the ``current'' trajectory for the Mental net, where the data was reversed and zero-padded whenever $t<\curtrajlen$. 
The teammate's future data in the current trajectory (a minimum of $30$ timesteps) was used for calculating various ground-truth targets against which to train the Prediction net. Namely, we used the following prediction targets: (a) the teammate's next step action, modeled as a discrete distribution, $\Delta(\actions)$; (b) the sign of each of its profile reward weights, $\rwdweights_i^k$, modeled as a distribution over $\{-1,0,1\}$; and (c) the successor representation (SR), corresponding to statistics of the teammate's future behavior. In particular, SR consisted of three types of data: (i) \textit{binary} features---whether the closest pot is empty, cooking or ready and similarly for the next closest pot (total $6$ features), encoded as Bernoulli distributions; (ii) \textit{categorical} features---relative frequency of the teammate carrying an onion, soup, dish, tomato or nothing (total $1$ feature), encoded as a discrete distribution; (iii) \textit{numeric} features---mean path distance to the teammate and $x$ and $y$ locations in the environment (total $3$ features), encoded as a multivariate Gaussian distribution. To obtain the SR means, we used discounted averaging over the teammate future data with a discount factor of $0.99$.

\subsubsection{Training and Losses}

The ToMnet architecture consisted of three networks as detailed in Sec.~\ref{Subsec:ToMnet}---a Character Net, a Mental Net and a Prediction Net. The Character and Mental Nets consisted of an LSTM layer of hidden size $64$ (dropout of $0.2$) that produces an initial embedding (final hidden state of the LSTM). The Character Net takes $\npast$ trajectory data each of length $\pasttrajlen=100$ as input, and to calculate the Character embedding, $\echar$, we first concatenated the LSTM's embedding with the observer's profile weights, sending the result through a dense/linear layer to obtain an embedding of size $\card{\echar}=8$. The Mental Net takes the current trajectory of length $\curtrajlen=10$ as input, where we masked out incomplete timesteps of a trajectory. To calculate the Mental embedding, $\emental$, we concatenated the LSTM's embedding with the profile's weights and $\echar$ before passing the result through a dense layer, resulting in an embedding of size $\card{\emental}=8$. 

Finally, to obtain the different predictions, we first concatenated both embeddings with the observer's profile weights and the current observation to obtain the prediction input. This is then sent through the different prediction heads to produce probabilistic predictions for the different targets, \ie, next action, profile, SR features detailed earlier. Each prediction head consists of a dense layer of size $64$ (dropout of $0.2$) with output dimension equaling the number of possible outputs for that prediction, namely, $|\actions|=6$ for the next action, $\card{\boldsymbol{\rwdweights}_j} \times 3 = 39$ for the profile sign prediction, $6$ for the SR binary features, $5$ for the SR categorical features, and $3$ for the SR numeric features. We used the negative log-likelihood (NLL) loss for the next action and profile predictions, binary cross-entropy loss for SR binary predictions, cross-entropy loss for SR categorical prediction, and a Gaussian NLL loss for SR numeric features' predictions.

For training the ToMnet, we used the $2\,000$ datapoints generated as explained above for each agent type, resulting in $1\,400$ datapoints, using a train-validation split of $80\%$-$20\%$ respectively. We used an Adam optimizer with weight decay of $5 \times 10^{-4}$. We trained the model for $2\,000$ epochs via mini-batch gradient descent with batch size of $128$ and learning rate of $5 \times 10^{-4}$, using an early stopping criterion based on a validation loss (maximum number of steps without improvement set to $10$ and a maximum tolerance of $0.01$).

\subsection{MADiff}%
\label{Subsec:AppMADiff}

\subsubsection{Trajectory Augmentation}

To train the MADiff component, we sampled joint trajectories similarly to how we trained the ToMnet, but augmented trajectories by computing the ToM embeddings for each timestep using the previously-trained ToMnet as explained in Sec.~\ref{Subsec:Training}. We used the same data parameters used to train the ToMnet, \ie, $\npast=4$ past trajectories of $\pasttrajlen=100$ steps each and $\curtrajlen=10$. We then generated trajectories to train the MADiff module by consecutively sampling from the augmented trajectories in a sliding window manner, using $\history=16$ steps to constrain trajectory generation using in-painting, and the subsequent $\horizon=64$ steps as the planning horizon. Trajectories were zero-padded at the beginning ($t<H)$) and end ($t>\trajlen-\horizon$) of each augmented trajectory during sampling.

\subsubsection{Training}
\label{Subsec:AppMADiffTrain}

For each sampled trajectory (length $\horizon + \history$), we augmented the observer agent's observations by including the one-hot encoding of the teammate's actions and then performed Cumulative Density Function (CDF) based normalization, corresponding to function $\obsnorm$ in Alg.~\ref{Alg:Planning}. For returns conditioning, we computed the discounted cumulative task reward in the original trajectory starting from the timestep in which the MADiff trajectory was sampled and then divided it by the maximum task reward that agents can receive, \ie $20$ in our case corresponding to a soup delivery. The reason for considering the whole episode instead of just the forward horizon for computing returns is that this way we denote the future potential of the sampled trajectory in achieving the task's goal (delivering soups), and not just what happens within the sample window itself. For the other conditioning variables, \ie observer profile, and character and mental embeddings, we normalized them using CDF similarly to observations. In summary, a single datapoint used for training the diffusion model consisted of a window of augmented observations, a return, an observer profile vector, and a character and a mental embedding. 

We trained the MADiff model with a history $\history=64$, horizon $\horizon=16$, and $\diffsteps=200$ diffusion steps, and an embedding and hidden dimensions of $128$ and $256$, respectively. We used epsilon noise prediction and loss computed as in Eq.~\ref{Eq:MADiffLoss}. We used dropout for conditions with a dropout rate of $0.25$ and conditional guidance with a factor of $\condguide=1.2$. We used $42\,000$ episodes, corresponding to $2\,000$ episodes per team (total of $21$ agent pairs). We used Adam optimizer with a batch size of $32$, learning rate of $2\times10^{-4}$ and trained each MADiff models for a total of $10^6$ training steps. For the different experiments reported in the paper, each MADiff model was trained with different conditioning variables, $\conditions{\traj{i}}$.

\subsubsection{Losses}

The integrated MADiff model, including the noise model $\paramdenoise$ and inverse dynamics model $\paraminvdyn$, is trained to optimize the loss $\loss{\paramdenoise,\paraminvdyn}$  \citep{zhu2024madiff}:
\begin{flalign}
\label{Eq:MADiffLoss}
    \loss{\paramdenoise,\paraminvdyn} &= \lossdiff + \lossinvdyn
    \\
    \lossdiff &= \mathbb{E}_{k,\jointtraj[0], \beta} \left[ \norm{\noise-\noisemodel(\jointobstraj[k],(1-\beta)\conditions{\traj{i}^0}+\beta \varnothing,k)}^2 \right]
    \\
    \lossinvdyn &= \mathbb{E}_{(\obs_i,\act_i,\obs'_i) \in \trajset} \left[\norm{\act_i - \invdyn(\obs_i,\obs'_i)}^2 \right]
\end{flalign}
where $i \sim \uniformset{\team}$ is the observer, $\jointtraj[0] \in \trajset$ is a joint trajectory sample where $\traj{i}^0$ is the observer's portion of it, $\beta \sim \bernoulli{p}$ balances conditional ($\conditions{\traj{i}^0}$) and unconditional ($\varnothing$) diffusion training, and $\noisemodel$ is the noise model optimizing the surrogate loss \citep{ho2020diff}, which estimates the noise $\noise \sim \normal{\boldsymbol{0}}{\boldsymbol{I}}$ added to sample $\jointtraj[0]$ at denoising step $k \sim \uniform{1}{\diffsteps}$.

\subsection{Experiments}%
\label{Subsec:AppExp}

Table~\ref{Table:ExpReplanning} presents the numeric data of the plots in Figs.~\ref{Fig:ResReplanTaskReward} and \ref{Fig:ResReplanIndivReward} of the replanning experiment.

\begin{table}[!ht]
    \centering
    \caption{Impact of various replanning schemes on the agents' cumulative task and individual rewards.}
    \label{Table:ExpReplanning}
    \begin{tabular}{l r @{\hspace{2pt}$\pm$\hspace{2pt}} r r @{\hspace{2pt}$\pm$\hspace{2pt}} r r @{\hspace{2pt}$\pm$\hspace{2pt}} r}
        \toprule
        \multicolumn{1}{l}{\textbf{Condition}}
        & \multicolumn{2}{c}{\textbf{Plan Count}}
        & \multicolumn{2}{c}{\textbf{Task Rwd}}
        & \multicolumn{2}{c}{\textbf{Indiv. Rwd}} \\
        \midrule

        \emph{Always} & $200.00$ & $0$ & $22.72$ & $2.95$ & $-1.08$ & $6.65$ \\
        \emph{10 Steps} & $23.00$ & $0$ & $18.24$ & $2.58$ & $-21.29$ & $9.06$ \\
        \emph{Horizon} & $4.00$ & $0$ & $13.76$ & $2.07$ & $-28.20$ & $9.15$ \\
        \emph{Dynamic} & $64.89$ & $3.82$ & $23.52$ & $2.89$ & $-1.74$ & $6.90$ \\

        \bottomrule
    \end{tabular}
\end{table}

\end{document}